\documentclass[12pt]{article}
\usepackage{amsmath,amssymb,epsfig,arydshln}

\usepackage{color}
\input{colordvi.tex}
\def\unit{{\relax{\rm 1\kern-.26em I}}}

\newcommand{\gsim}{\gtrsim}

\def\6#1{{\underline{#1}}}
\def\m6#1{{\underline{#1}\,}}

\newdimen\Tdim
\def\ispan{{\setbox0=\hbox{i}%
\Tdim\ht0\advance\Tdim\dp0\rule[-\dp0]{0pt}{\Tdim}}}
\def\jspan{{\setbox0=\hbox{j}%
\Tdim\ht0\advance\Tdim\dp0\rule[-\dp0]{0pt}{\Tdim}}}
\def\Tspan#1{{\setbox0=\hbox{#1}%
\Tdim\ht0\advance\Tdim\dp0\advance\Tdim.55ex\rule[-\dp0]{0pt}{\Tdim}\box0}}

\def\be{\begin{eqnarray}}
\def\ben{\begin{eqnarray*}}
\def\ee{\end{eqnarray}}
\def\een{\end{eqnarray*}}
\def\Tr{{\rm Tr}}
\def\Lag{\mathcal{L}}
\def\p{\partial}

\def\D{\mathcal{D}}

\def\=:{=\hspace{-.7em}\raisebox{1.1ex}{.}\hspace{.1em}\raisebox{-0.2ex}{.} }

\newcommand{\NF}{N_{\rm F}}
\newcommand{\NC}{N_{\rm C}}

\def\N{{\cal N}}

\newcommand {\beq}{\begin{eqnarray}}
\newcommand {\eeq}{\end{eqnarray}}
\newcommand {\non}{\nonumber\\}

\makeatletter

\renewcommand\section{\@startsection {section}{1}{\z@}%
                                   {-3.5ex \@plus -1ex \@minus -.2ex}%
                                   {2.3ex \@plus.2ex}%
                                   {\normalfont\large\bfseries}}

\renewcommand\subsection{\@startsection{subsection}{2}{\z@}%
                                     {-3.25ex\@plus -1ex \@minus -.2ex}%
                                     {1.5ex \@plus .2ex}%
                                     {\normalfont\normalsize\bfseries}}

\newcount\hour \newcount\minute
\hour=\time \divide \hour by 60
\minute=\time
\def\now{%
\ifnum \hour<13
  \ifnum \hour=0 \advance \hour by 12 \number\hour:\else \number\hour:\fi%
     \ifnum \minute<10 0\fi%
     \number\minute%
\ A.M.%
\else \advance \hour by -12 \number\hour:%
  \ifnum \minute<10 0\fi%
  \number\minute%
  \ P.M.%
\fi%
}

\makeatother

\begin{document}

\baselineskip=18pt  
\numberwithin{equation}{section}  
\allowdisplaybreaks  



%
%


\thispagestyle{empty}

\vspace*{-2cm}
\begin{flushright}
{\tt YGHP-14-03}
\end{flushright}

\begin{flushright}
\end{flushright}

\begin{center}

\vspace{-.5cm}

\vspace{0.5cm}
{\bf\Large Dyonic non-Abelian vortex strings\\ in\\ supersymmetric and non-supersymmetric theories}
\\\ \\
{\bf\large--- tensions and higher derivative corrections ---}
\vspace*{1.5cm}

{\bf
Minoru Eto$^a$\footnote{\it e-mail address:
meto(at)sci.kj.yamagata-u.ac.jp}
and Yoshihide Murakami$^a$\footnote{\it e-mail address:
cosmoboe(at)gmail.com}}
\vspace*{0.5cm}
 
\vspace*{0.5cm}

$^a$ {\it {Department of Physics, Yamagata University, Kojirakawa-machi 1-4-12, Yamagata, Yamagata 990-8560, Japan}}\\

\end{center}

\vspace{1cm} \centerline{\bf Abstract} \vspace*{0.5cm}

Dyonic non-Abelian local/semi-global vortex strings are studied in detail in supersymmetric/non-supersymmetric
Yang-Mills-Higgs theories. While the BPS tension formula is known to be the same as that for the 
BPS dyonic instanton,  we find that 
the non-BPS tension formula is approximated very well by the well-known tension formula of 
the BPS dyon. We show that this mysterious tension formula for the dyonic non-BPS vortex stings
can be understood from the perspective of
a low energy effective field theory. Furthermore, 
we propose an efficient method to obtain an effective  theory of a single vortex string, which
includes not only lower derivative terms but also all order derivative corrections
by making use of the tension formula.
We also find a novel dyonic vortex string whose internal orientation vectors rotate in time and
spiral along the string axis.

\newpage
\setcounter{page}{1} 





\section{Introduction}

When a non-trivial symmetry is spontaneously broken, it is possible that topological
solitons, which are non-perturbative objects, appear in spectra. They are non-dissipative lumps of energy and
behave as if they are isolated particles. They play important roles in quantum field theories. 
For instance, instantons are necessary  
for understanding non-perturbative structure of non-Abelian gauge theories, and magnetic monopoles
may solve the long standing problem of color confinement in QCD via the so-called dual Meissner effect~\cite{Nambu:1974zg,'tHooft:1981ht,Mandelstam:1974pi}, 
and so on. 

There exist topological solitons, the so-called Bogomol'nyi-Prasad-Sommerfield (BPS)~\cite{Bogomolny:1975de,Prasad:1975kr}
solitons, 
which have several special features and
are deeply related to supersymmetry. Although an inhomogeneous field configuration generally  breaks
supersymmetry, the BPS solitons  preserve 
a part of it.
Furthermore, thanks to the supersymmetry, net interactions among multiple BPS solitons vanish, so that  
their masses are proportional to topological winding numbers.
While many kinds of BPS solitons are characterized by only topological charges,
there also exist extended BPS solitons which are also characterized by
conserved charges for some local/global symmetries.
They are called dyonic solitons or Q-solitons in the literature.
Several examples are in order: dyonic instantons in $1+4$ dimensions~\cite{Lambert:1999ua}, dyons
in $1+3$ dimensions~\cite{Julia:1975ff}, Q-lumps in $1+2$ dimensions~\cite{Leese:1991hr}, and Q-kinks in $1+1$ dimensions~\cite{Abraham:1992vb,Abraham:1992qv}. The charged semi-local Abelian vortices in $1+2$ dimensions~\cite{Abraham:1992hv} are
also one of the dyonic BPS solitons. As explained in Appendix, these dyonic BPS solitons have
special mass formulae, and they can be classified into two types: the dyonic-instanton
type and the dyon type. The former has the BPS mass formula
\beq
M_{\text{d-inst}} = |T| + |Q|,
\label{eq:instanton-type}
\eeq
while the latter has another formula
\beq
M_{\rm dyon} = \sqrt{T^2 + Q^2},
\label{eq:dyon-type}
\eeq
where $T$ and $Q$ stand for a topological charge and an electric (Noether) charge, respectively.
The dyonic instantons, the Q-lumps, and the charged semi-local Abelian vortices belong to the
dyonic-instanton type while the dyons and the Q-kinks are of the dyon type.
Common features of these solitons are that firstly they are all BPS states, 
and secondly spatial co-dimensions of the
solitons of the dyonic-instanton type are even while those of the dyon-type solitons are odd.
Thus, we naively guess  that {\it all dyonic solitons belonging to the dyonic-instanton (dyon) type are 
BPS and their spatial co-dimensions are even (odd).}

Let us examine this conjecture by 
another dyonic soliton which was found relatively recently compared to the other dyonic
solitons listed above. It is called dyonic non-Abelian local vortex (DNALV) in
$2+1$ dimensions~\cite{Collie:2008za}. It is a dyonic extension of
the so-called non-Abelian local vortex (NALV).
A non-Abelian local vortex string (NALVS)~\cite{Hanany:2003hp,Auzzi:2003fs,Shifman:2004dr,Hanany:2004ea}
in $3+1$ dimensional 
Yang-Mills-Higgs theories is a natural extension 
of the Nielsen-Olsen vortex string in Abelian-Higgs theories, 
which is also called the Abelian local vortex string (ALVS).
NALVS carries a genuine non-Abelian magnetic flux. In other words, they have non-Abelian
moduli parameters that are points on a compact manifold, typically $\mathbb{C}P^{\NC-1}$~\cite{Hanany:2003hp,Auzzi:2003fs,Shifman:2004dr,Hanany:2004ea}.
So far, lots of works have been done for revealing many roles by the non-Abelian vortex strings in supersymmetric
theories, see the review articles~\cite{Tong:2005un,Eto:2006pg,Konishi:2007dn,Shifman:2007ce} and references therein, and see also recent studies~\cite{Shifman:2013oia,Eto:2013hoa,Bolokhov:2013bea,Monin:2013mxa,Evslin:2013wka,Kobayashi:2013axa,Shifman:2014lba,Shifman:2014oqa,Chatterjee:2014rqa,Arai:2014hda,Bolognesi:2014saa,Nitta:2013wca}. 
While much of them have been focused on static BPS configurations, few studies
concentrated on
a time-dependent (stationary) configuration, namely dyonic non-Abelian local vortex string (DNALVS). 
As will be explained in Sec.~\ref{sec:2}, in a certain condition it is a BPS object of co-dimensions two.
The BPS mass formula was found to be 
of the dyonic-instanton type~\cite{Collie:2008za}. Hence, DNALVS makes us further believe
that our conjecture is correct.

The aim of this paper is to study dyonic non-Abelian vortex strings in more detail.
Firstly, we will focus on DNALVS
in supersymmetric $U(\NC)$ Yang-Mills-Higgs model in $3+1$ dimensions.
Among the known BPS dyonic solitons listed above,
DNALVSs have a distinct feature that they can continuously change from BPS 
to non-BPS by increasing Noether charges~\cite{Collie:2008za}.
While the BPS DNALV in $2+1$ dimensions was studied in Ref.~\cite{Collie:2008za}, 
non-BPS one with a large
Noether charge has not been studied in detail. Especially, a tension (mass per unit length)
formula for the non-BPS state
is not clarified so far. Generally speaking, there are no reasons for expecting any simple mass formulae
like Eqs.~(\ref{eq:instanton-type}) and (\ref{eq:dyon-type}) for non-BPS solitons.
Nevertheless, we find an interesting result that 
tensions of the non-BPS DNALVSs are reproduced pretty well by
the dyon-type tension formula. However, unlike BPS cases, it is not so easy to derive some exact results for
non-BPS states. Actually, we numerically verify that DNALVS appears to obey the dyon-type formula, so that
it is hard for us to conclude that the dyon-type formula is an exact result.
It is a little bit pity that we will proof that the dyon-type tension formula cannot be exact. However, the fact that
the dyon-type formula is just an approximation but accurately reproduces the tensions is still surprising.
Thus, DNALVS changes the type according to the magnitude of the Noether charge.
For a small Noether charge, it remains BPS and belongs to the dyonic-instanton type. On the other hand,
for a large Noether charge, it becomes non-BPS and changes to the approximate dyon type.
To the best of our knowledge, no such solitons having this feature have been known in the literature.

In order to investigate such novel solitons further, we will next study a dyonic extension of a similar non-Abelian
vortex string in a non-supersymmetric $SU(\NC)$ Yang-Mills-Higgs model. 
Ungauging the overall $U(1)$ gauge symmetry of $U(\NC)$ 
makes the non-Abelian vortex string a sort of global vortex string whose tension  
is logarithmically divergent. 
They were first found in the color-flavor locked phase of color superconductor in the high density QCD, and
are sometimes called the semi-superfluid vortex strings~\cite{Balachandran:2005ev,Nakano:2007dr,Nakano:2008dc,Eto:2009kg,Eto:2009bh,Eto:2009tr,Eto:2011mk,Eto:2013hoa}. In this work, superfluidity is not relevant,
so we christen them non-Abelian semi-global  
vortex strings (NASGVSs) in this paper. 
It is an important feature that NASGVS has a well-squeezed flux tube of 
non-Abelian magnetic fields inside its logarithmically expanded core of the scalar fields.
This implies that they have normalizable non-Abelian zero modes, typically $\mathbb{C}P^{\NC-1}$.
Furthermore, they are non-BPS solitons even when they carry no Noether charges.
We numerically find a non-BPS dyonic extension of NASGVS by taking a rotation inside
the internal moduli space $\mathbb{C}P^{\NC-1}$ into account.
While its tension is still logarithmically divergent, we numerically find that a difference in tensions of the dyonic one
and non-dyonic one remains finite. Remarkably, the tension formula again is of the approximate 
dyon type.


Furthermore, we will try to go beyond the numerical and accordingly approximate tension formula for
the non-BPS DNALVS. For that purpose, we will consider DNALVS in $2+1$ dimensions, and 
will make use of the low energy effective theory
on the vortex string world volume. Apart from the center of mass of DNALVS, the low energy effective
theory is $0+1$ dimensional $\mathbb{C}P^{\NC-1}$ non-linear sigma model~\cite{Tong:2005un,Eto:2006pg,Konishi:2007dn,Shifman:2007ce}.
So far, much of works have studied the effective theory with quadratic derivative terms, which
is a lowest approximation taking only massless degrees of freedom into account. 
In general, finding higher derivative corrections is not a very easy task. Indeed only the next leading 
order, namely a quartic derivative correction, was obtained~\cite{Eto:2012qda}.
For our purpose, this quartic derivative term will play an important role. We will combine it 
with the numerical result for the tension of DNALVS which is obtained by solving the equations of motion
for not the effective theory but the original $2+1$ dimensional theory.
As a result, we will find a plausible low energy effective Lagrangian including 
higher derivative corrections to all order. This Lagrangian has only one parameter which
can be determined by the next leading order effective Lagrangian.
From the effective Lagrangian, we can derive analytic expressions of the tension
and the Noether charge. Hence, we will find an implicit relation between
them, which reproduces the numerical results very well.
The method used here for deriving the low energy effective theory including all order higher derivative
corrections may be possible to apply for other topological solitons. 
As mentioned above, derivation of higher derivative corrections is usually very involved.
Therefore, our method may offer an efficient way.

This paper is organized as follows. In Sec.~\ref{sec:2}, we will mainly explain in detail 
the BPS (D)NALVS in the supersymmetric $U(\NC)$ Yang-Mills-Higgs theory. The most of this section will be
a review but the results in Sec.~\ref{sec:lorentz boost} are new. The main results of this work
will be shown in subsequent Secs.~\ref{sec:non_BPS_dyonic_U2_vortex} -- \ref{sec:massless_dyonic_SGV}.
Sec.~\ref{sec:non_BPS_dyonic_U2_vortex} will be devoted to the non-BPS DNALVSs.
We will numerically solve the full equations of motion and find the tension formula of the approximate
dyon type. In Sec.~\ref{sec:higher_derivatives}, 
we will derive a low energy effective Lagrangian of DNALVS including higher derivative
corrections to all order. Making use of it, we will find an implicit relation between
the tension and the Noether charge, which improves the approximate dyon-type tension formula.
In Sec.~\ref{sec:massless_dyonic_SGV}, we will study
the non-BPS DNASGVSs in non-supersymmetric $SU(\NC)$ Yang-Mills-Higgs
theory. We will conclude this work in Sec.~\ref{sec:conc}. Several BPS dyonic solitons in various models 
are reviewed in the Appendix.

\section{BPS dyonic non-Abelian local vortex strings}
\label{sec:2}

In this section we briefly review on the BPS NALVSs~\cite{Tong:2005un,Eto:2006pg,Konishi:2007dn,Shifman:2007ce} and their dyonic extensions~\cite{Collie:2008za}
in $\N=2$ supersymmetric $U(\NC)$ gauge theories.
While Secs.~\ref{sec:bps NA vortex} -- \ref{sec:dnalvs_eff_th} are almost reviews,
Sec.~\ref{sec:lorentz boost} includes new results.

\subsection{Non-Abelian vortex string}
\label{sec:bps NA vortex}

A bosonic part of ${\cal N}=2$ SUSY Lagrangian in $3+1$ dimensions is given by
\beq
\Lag_{{\cal N}=2} &=& \Tr\bigg[
-\frac{1}{2g^2}F_{\mu\nu}F^{\mu\nu} + \frac{1}{g^2}\D_\mu \Sigma (\D^\mu\Sigma)^\dagger
+ \D_\mu H^i (\D^\mu H_i)^\dagger  \non
&-& \frac{g^2}{4}\left((\sigma_a)^i{}_jH^iH_j^\dagger - c_a{\bf 1}_{\NC}\right)^2
- (\Sigma H^i - H^iM)(\Sigma H^i - H^iM)^\dagger
\bigg].
\label{eq:lag_U(N)}
\eeq
Here, $W_\mu$ and $\Sigma$ belong to a $U(\NC)$ vector multiplet. An $\NC$ by $\NF$ matrix field
$H^{i}$ ($i=1,2$ is $SU(2)_{\rm R}$ index), the lowest component of hypermultiplet, is in the fundamental representation of $U(\NC)$ gauge 
group.
Covariant derivatives are given by
\beq
F_{\mu\nu} &=& \p_\mu W_\nu - \p_\nu W_\mu + i \left[W_\mu,W_\nu\right],\\
\D_\mu \Sigma &=& \p_\mu \Sigma + i \left[W_\mu,\Sigma\right],\\
\D_\mu H^i &=& \p_\mu H^i + i W_\mu H^i.
\eeq
We will use a matrix notation for $W_\mu = W_\mu^aT^a$ and $\Sigma = \Sigma^aT^a$ 
($\Tr[T^aT^b] = \delta^{ab}/2$). The Lagrangian has three coupling constants:
$g$ stands for a $U(\NC)$ gauge coupling constant, $c_a$ ($a=1,2,3$) is the $SU(2)_R$ triplet
of the Fayet-Iliopoulos parameter (in what follows, we set $c_a = (0,0, v^2)$, ($v>0$), 
by using $SU(2)_R$ transformation), and 
$M$ is a diagonal mass matrix as $M = {\rm diag}
(m_1,m_2,\cdots,m_{\NF})$. In general, the elements $m_i$ are complex variables but we impose them to be 
real valued, and we order them as $m_i \ge m_{i+1}$ without loss of generality. Furthermore, 
the mass matrix $M$
can be always set to be traceless ($\sum_i m_i = 0$) by shifting $\Sigma$ by a constant. 
Thus, the potential becomes
\beq
V &=& \Tr\bigg[\frac{g^2}{4}\left(H^1H_1^\dagger + H^2H_2^\dagger - v^2{\bf 1}_{\NC}\right)^2
+ g^2 (H^1H_2^\dagger)(H^1H_2^\dagger)^\dagger\non
&&+\ (\Sigma H^i - H^iM)(\Sigma H^i - H^iM)^\dagger
\bigg].
\eeq
In the massless case $M=0$,
the model has not only $U(\NC)$ gauge symmetry but also $SU(\NF)$ flavor symmetry, which
act on the fields as
\beq
H^i \to U_{\rm C} H^i U_{\rm F},\quad
\Sigma \to U_{\rm C} \Sigma U_{\rm C}^\dagger,\qquad
U_{\rm C} \in U(\NC),\quad
U_{\rm F} \in SU(\NF).
\eeq
In general case with $M\neq 0$, the flavor symmetry $SU(\NF)$
reduces to a subgroup which commutes with the mass matrix $M$. The minimal flavor symmetry
is the maximum torus $U(1)^{\NF-1}$ in the case that all the elements of the mass matrix are not degenerate,
namely $m_i > m_{i+1}$.

It is well known that the above model possesses rich topological excitations,
for example, magnetic monopoles, vortex strings and domain walls as half and/or quarter BPS states~\cite{Shifman:2002jm,Shifman:2003uh,Isozumi:2004vg,Sakai:2005sp}.
In Sec.~\ref{sec:2} and \ref{sec:non_BPS_dyonic_U2_vortex}, 
we will put our focus onto  NALVSs which appear in
the case of $\NC = \NF$ with $M=0$.
So, we will concentrate on the case $\NC=\NF$ hereafter. 
Before going to the vortex strings, we explain the vacuum of the model.
A supersymmetric vacuum is uniquely determined up to gauge transformation as 
\beq
H^1 = v{\bf 1}_{\NC},\quad 
H^2 = 0,\quad 
\Sigma = 0.
\eeq
The vacuum is invariant under the color-flavor locked (CFL) {\it global} symmetry
\beq
(H^1, H^2,\Sigma) \to U_{\rm C+F} (v{\bf 1}_{\NC},0,0) U_{\rm C+F}^\dagger ,\qquad
U_{\rm C+F} \in SU(N)_{\rm C+F}.
\eeq
Thus, this vacuum is in the so-called CFL phase. Note that the vacuum is unique and all the massive
fields have the same mass thanks to the $\N=2$ supersymmetry
\beq
\mu^2 = g^2 v^2.
\eeq

Next, we consider the BPS NALVSs.
In the following, 
we will set $H^2=0$ because $H^2$ will be irrelevant to the subsequent discussions.  
This is, of course, consistent with the equations of motion. 
Furthermore, we will use simple notation $H^1 \to H$ in the rest of the paper.
Then, the equations of motion
for the non-trivial fields $H$, $W_\mu$ and $\Sigma$ are given by
\beq
- \frac{1}{g^2}\D_\mu F^{\mu\nu} &=& \frac{i}{g^2}\left\{\left[\Sigma,(\D^\nu\Sigma)^\dagger\right]
-\left[\D^\nu\Sigma,\Sigma^\dagger\right]
\right\} + i \left\{H(\D^\nu H)^\dagger - (\D^\nu H) H^\dagger\right\},
\label{eq:eom_W}\\
-\frac{1}{g^2}\D_\mu\D^\mu \Sigma &=& \left(\Sigma H - H M\right) H^\dagger,
\label{eq:eom_Sigma}\\
\D_\mu\D^\mu H &=& - \frac{g^2}{2}\left(HH^\dagger - v^2 {\bf 1}_N\right)H
+ \left(\Sigma H - H M\right)M - \Sigma^\dagger \left(\Sigma H - H M\right).
\label{eq:eom_H}
\eeq
Here  we retain the generic mass matrix $M$.
These are complicated second order partial differential equations, so that it is not easy to solve them
in practice. However, for BPS states, 
one does not need to solve the equations of motion.
Instead, we solve BPS equations. In this work, we are interested in a BPS vortex string. For simplicity, 
we assume that it is a straight string perpendicular to the $x^1x^2$ plane, so we will set $W_\alpha =0$
and $\p_\alpha =0$ with $\alpha =0,3$. Furthermore, we will set $\Sigma = 0$ which solves 
Eq.~(\ref{eq:eom_Sigma}) when $M=0$.
Now we are ready to derive the BPS equations which can be derived  through a standard
Bogomol'nyi technique for the Hamiltonian. 
Namely, the Hamiltonian per unit length can be cast into the following perfect square form
\beq
H &=& \int dx^1dx^2\ \Tr\bigg[
\frac{1}{g^2}(F_{12})^2  
+ \D_m H (\D_m H)^\dagger 
+ \frac{g^2}{4}\left(HH^\dagger  - v^2{\bf 1}_{\NC}\right)^2
\bigg] \non
&=& \int dx^1dx^2\ \Tr\bigg[
\frac{1}{g^2}\left(F_{12} - \frac{g^2}{2} \left(HH^\dagger - v^2{\bf 1}_{\NC}\right)\right)^2
+ (\D_1 H + i \D_2 H)(\D_1 H + i \D_2 H)^\dagger\non
&&- v^2 F_{12} + \p_m {\cal J}_m
\bigg] \non
&\ge& -v^2 \int dx^1dx^2\ \Tr[F_{12}],
\eeq
where $m=1,2$ and we define ${\cal J}_m \equiv i \epsilon_{mn} H(\D_n H)^\dagger$ which rapidly goes to
zero at the boundary of $x^1x^2$ plane since $\D_m H = 0$ in the vacuum. 
The Bogomol'nyi bound is saturated when
\beq
(\D_1+i\D_2) H = 0,\quad F_{12} = \frac{g^2}{2}\left(HH^\dagger - v^2 {\bf 1}_{\NC}\right),
\label{eq:NA_vor_BPS_eq}
\eeq
are satisfied, 
and the tension of the BPS NALVS  reads
\beq
T  = - 2\pi v^2 \int d^2x\ \Tr[F_{12}] = 2\pi v^2 k,\qquad k \in \mathbb Z_{>0}.
\eeq
One can easily find anti-BPS solitons for a negative integer $k$ 
by appropriately adjusting signs in the BPS equations.
It is obvious that  any solutions of the BPS equations solve the equations of motion (\ref{eq:eom_W}) and
(\ref{eq:eom_Sigma})
because the BPS states saturate the energy bound from below. 

\begin{figure}[t]
\begin{center}
\includegraphics[width=12cm]{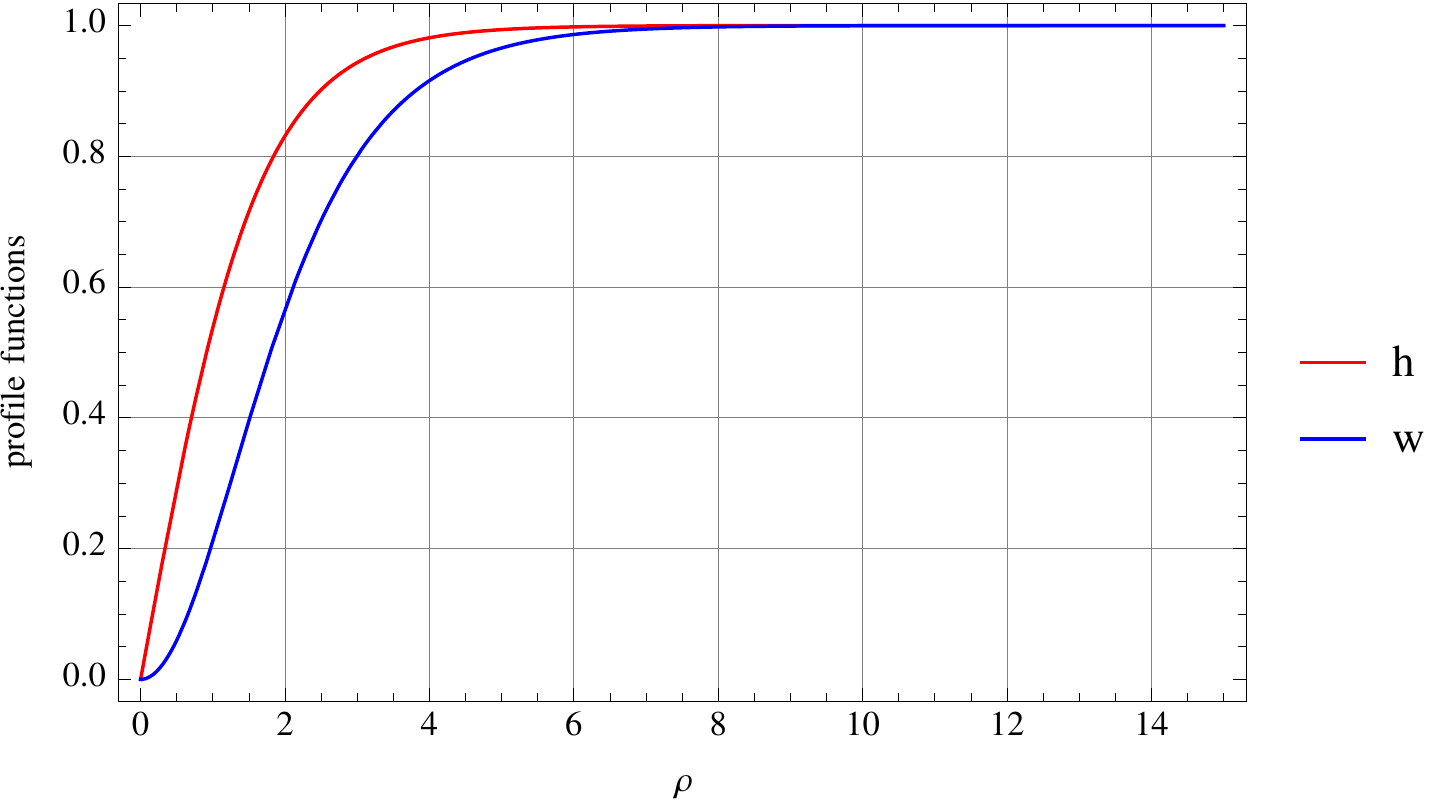}
\caption{The profile functions $h(\rho)$ and $w(\rho)$ for the minimally winding NALVS.}
\label{fig:BPS_NALVS}
\end{center}
\end{figure}

The BPS equations are the first order partial differential equations, so solving them is much easier
compared to the equations of motion. In order to further simplify the BPS equations, let us use the symmetry.  
A straight string is axially symmetric, so one can make a natural Ansatz for minimally winding ($k=1$) NALVS
\beq
H(x^1,x^2) &=& v h(r)e^{i\theta}{\rm diag.}\left(
1,0,\cdots,0\right),
\label{eq:NA_BPS_ansatz_1}\\
W_1(x^1,x^2) + i W_2(x^1,x^2) &=& 
- i  \frac{w(r)e^{i\theta}}{r}
{\rm diag.}\left(
1,0,\cdots,0\right),
\label{eq:NA_BPS_ansatz_2}
\eeq
with $x^1+ix^2 = r e^{i\theta}$. The magnetic field $F_{12}$ is expressed by
\beq
F_{12} = - \frac{w'(r)}{r}{\rm diag.}
\left(
1,0,\cdots,0\right).
\eeq
Plugging these into Eq.~(\ref{eq:NA_vor_BPS_eq}), we are lead to
first order ordinary differential equations for $h(\rho)$ and $w(\rho)$ with respect to a dimensionless coordinate
\beq
\rho \equiv \mu r,
\eeq
as 
\beq
h'(\rho) + \frac{h(\rho)}{\rho}(w(\rho)-1) = 0,\quad 
- \frac{w'(\rho)}{\rho} = \frac{1}{2}(h(\rho)^2 -1),
\label{eq:NA_BPS_axial}
\eeq
where the prime stands for a derivative by $\rho$.
These should be solved with the boundary conditions
\beq
h(0) = 0,\quad w(0) = 0,\qquad
h(\infty) = 1,\quad w(\infty) = 1,
\eeq
which guarantees a non-trivial topological charge ($k=1$)
\beq
- \frac{1}{2\pi} \int d^2x\ \Tr[F_{12}] = - \int_0^\infty dr\ r \left(-\frac{w'}{r}\right) = 1.
\eeq
Unfortunately, no analytic solutions for Eq.~(\ref{eq:NA_BPS_axial}) have been known, so 
we numerically solve them. A numerical solution is shown in Fig.~\ref{fig:BPS_NALVS}.

Note that the BPS equations in Eq.~(\ref{eq:NA_BPS_axial}) are exactly identical to
those of the Nielsen-Olesen vortex string in the BPS limit of the Abelian-Higgs model, see for example Ref.~\cite{Manton:2004tk}.
In this sense, the $k=1$ BPS NALVS can be thought of as a merely embedding solution of 
the BPS Nielsen-Olesen vortex string into the $\NC\times\NC$ matrices in the $U(\NC)$ non-Abelian gauge theory.
This realization immediately leads us to non-Abelian moduli parameters corresponding to degrees of
freedom for the embedding.
A key ingredient is
the CFL symmetry $SU(\NC)_{\rm C+F}$ of the vacuum state. The single NALVS given
in Eqs.~(\ref{eq:NA_BPS_ansatz_1}) and (\ref{eq:NA_BPS_ansatz_2}) spontaneously breaks it
down to $U(1)_{\rm C+F} \times SU(\NC-1)_{\rm C+F}$. 
Thus there are infinitely degenerate states whose moduli
space is given by
\beq
{\cal M}_{k=1} = \mathbb{C} \times \frac{SU(\NC)_{\rm C+F}}{U(1)_{\rm C+F} \times SU(\NC-1)_{\rm C+F}}
\simeq \mathbb{C} \times \mathbb{C}P^{\NC-1}.
\eeq
The first factor $\mathbb{C}$ corresponds to the position of the vortex string on the $x^1x^2$ plane,
while the second factor $\mathbb{C}P^{\NC-1}$ is genuine non-Abelian moduli which is the so-called
orientational moduli, see for example 
Ref.~\cite{Tong:2005un,Eto:2006pg,Konishi:2007dn,Shifman:2007ce}.

\subsection{A dyonic extension}
\label{sec:dyonic ext}

Let us now extend the BPS NALVSs in the previous subsection, which are static
configurations, to dyonic solutions~\cite{Collie:2008za} by allowing time dependence.
In order to remain solutions to be BPS states, we now need to include the non-trivial
adjoint scalar field $\Sigma$ and the generic mass matrix $M$. So we restore $\Sigma$ and $W_0$
while set  $H^2$ to zero as before. We still keep $\NC=\NF$, and
the flavor symmetry is the maximum torus $U(1)^{\NF-1}$.
We begin with the Bogomol'nyi completion of the Hamiltonian
\beq
H &=& \int dx^1dx^2\ \Tr\bigg[
\frac{1}{g^2}\left\{(F_{01})^2 + (F_{02})^2 + (F_{03})^2 + (F_{12})^2 + (F_{23})^2 + (F_{31})^2\right\}  \non
&& + \frac{1}{g^2}\left\{
\D_0\Sigma (\D_0\Sigma)^\dagger + \D_3\Sigma (\D_3\Sigma)^\dagger 
+ \D_m\Sigma (\D_m \Sigma)^\dagger\right\} \non
&& + \D_0 H (\D_0 H)^\dagger + \D_3 H (\D_3 H)^\dagger + \D_m H (\D_m H)^\dagger \non
&& + \frac{g^2}{4}\left(HH^\dagger  - v^2{\bf 1}_{\NC}\right)^2
+ (\Sigma H - H M)(\Sigma H - H M)^\dagger
\bigg] \non
&=& \int dx^1dx^2\ \Tr\bigg[
\frac{1}{g^2}\left(F_{12} - \frac{g^2}{2} \left(HH^\dagger - v^2{\bf 1}_{\NC}\right)\right)^2
+ (\D_1 H + i \D_2 H)(\D_1 H + i \D_2 H)^\dagger\non
&& + \frac{1}{g^2}(F_{0m} \mp \D_m \Sigma)(F_{0m} \mp \D_m \Sigma)^\dagger 
+ \left(\D_0 H \pm i (\Sigma H - HM)\right)\left(\D_0 H \pm i (\Sigma H - HM)\right)^\dagger\non
&& + \frac{1}{g^2}\left\{
(\D_0\Sigma)(\D_0\Sigma)^\dagger + (\D_3\Sigma)(\D_3\Sigma)^\dagger + (F_{03})^2 
+ (F_{23})^2 + (F_{31})^2
+ \D_3 H (\D_3H)^\dagger
\right\} \non
&& - v^2 \Tr [F_{12}] + \p_m {\cal J}_m \pm \frac{1}{g^2}\p_m\left((\Sigma + \Sigma^\dagger)F_{0m}\right)
\mp i \left\{(\D_0H) M H^\dagger - H M (\D_0H)^\dagger\right\} \non
&& \pm \Sigma \left(-\frac{1}{g^2}\D_m F_{0m} - i H(\D_0H)^\dagger\right)
\pm \Sigma^\dagger \left(-\frac{1}{g^2}\D_m F_{0m} + i (\D_0H) H^\dagger\right)\bigg] \non
&\ge& \int dx^1dx^2\ \Tr\left[-v^2 F_{12} \right] + \left|\int dx^1dx^2\ \Tr_{\rm F} \left[
M J_{\rm F}^0 \right] \right|,
\label{eq:Bogo}
\eeq
where vanishing of the last line of the second equality of Eq.~(\ref{eq:Bogo}) will be explained at the end 
of this subsection. 
We have defined the $SU(\NC)_{\rm F}$ flavor current by
\beq
J_{\rm F}^\mu \equiv - i \left( H^\dagger \D^\mu H - (\D^\mu H)^\dagger H\right).
\label{eq:flavor_current}
\eeq
Note that among elements of $J_{\rm F}^\mu$  only the elements which 
commute with the mass matrix $M$ are conserved. The conserved charge per unit length is defined by
\beq
Q^a = 2 \int dx^1dx^2\ \Tr\left[J_{\rm F}^0 T^a\right],\quad \text{for}\quad \left[T^a,M\right] = 0.
\label{eq:Q_SU(2)}
\eeq
Thus, the tension of the BPS DNALVSs reads 
\beq
T = 2\pi v^2 k + \left|\Tr[MQ]\right|,
\label{eq:BPS_dv_mass}
\eeq
with $Q = Q^aT^a$.
As is mentioned in the Introduction, this is the dyonic-instanton type.
Note that there is a maximum value for the second term $|\Tr[MQ]|$, namely it cannot be arbitrary large.
A reason for that will be given in the subsequent section \ref{sec:low energy}.

Eq.~(\ref{eq:Bogo}) also includes the electric charge density
\beq
{\cal Q}_{\rm e} = \frac{1}{g^2}\p_m \Tr\left[(\Sigma + \Sigma^\dagger) F_{0m}\right].
\eeq
Hence, the dyonic vortex is electrically charged object but the net electrical charge is zero,
$\int dx^1dx^2\ {\cal Q}_{\rm e} = 0$,
since it is screened in the Higgs phase (a superconductor).

The energy bound is saturated when the following first order equations are satisfied
\beq
(\D_1+i\D_2) H = 0,\quad F_{12} = \frac{g^2}{2}\left(HH^\dagger - v^2 {\bf 1}_{\NC}\right),
\label{eq:NA_vor_BPS_eq_again}\qquad\quad\\
F_{23} = F_{31} = F_{03} = 0,\quad \D_3 \Sigma = 0,\quad \D_3 H = 0,\qquad\quad
\label{eq:NA_vor_BPS_eq_new_3}\\
F_{0m} \mp \D_m\Sigma = 0,\quad
\D_0 H \pm i (\Sigma H - H M) = 0,\quad \D_0 \Sigma = 0.
\label{eq:NA_vor_BPS_eq_new}
\eeq
These first order equations can be also derived from a 1/4 BPS condition in $\N=2$ SUSY Yang-Mills-Higgs
theories~\cite{Eto:2005sw}.
Eq.~(\ref{eq:NA_vor_BPS_eq_again}) is exactly the same as Eq.~(\ref{eq:NA_vor_BPS_eq}).
Therefore, $x^1$ and $x^2$ dependence of the solution for $H$ and $W_{1,2}$ are 
identical to those obtained  by solving Eq.~(\ref{eq:NA_BPS_axial}).
Eq.~(\ref{eq:NA_vor_BPS_eq_new_3}) is trivially satisfied by $\p_3 = 0$ and $W_3 = 0$.
In order to solve the first equation in Eq.~(\ref{eq:NA_vor_BPS_eq_new}), we fix the gauge by
\beq
\Sigma(x^1,x^2) = \mp W_0(x^1,x^2).
\eeq
Note that this also solves $\D_0\Sigma = 0$.
Then the second equation of Eq.~(\ref{eq:NA_vor_BPS_eq_new}) becomes
\beq
\p_0 H = \pm i H M.
\eeq
This can be easily solved by
\beq
H(t,x^1,x^2) = H_\star(x^1,x^2)e^{\pm iMt},
\label{eq:H_dyonic}
\eeq
where $H_\star(x^1,x^2)$ can be any of  {\it static} BPS solutions for 
Eq.~(\ref{eq:NA_vor_BPS_eq_again}).
Finally, we should determine $W_0(x^1,x^2)$ by solving the Gauss law
\beq
-\frac{2}{g^2}\D_mF^{m0} - \frac{i}{g^2}\left\{
\left[\Sigma, (\D^0\Sigma)^\dagger\right] - \left[\D^0\Sigma,\Sigma^\dagger\right]\right\}
-i \left\{H(\D^0H)^\dagger - (\D^0H)H^\dagger\right\} = 0.
\label{eq:Gauss}
\eeq 
By using Eq.~(\ref{eq:NA_vor_BPS_eq_new}), this can be rewritten as
\beq
\frac{2}{g^2}\D_m\D^m\Sigma +
H_\star\left(H_\star^\dagger \Sigma - MH_\star^\dagger\right) + (\Sigma H_\star - H_\star M)H_\star^\dagger = 0
\label{eq:eq_Sigma}
\eeq
Note that this is consistent with the equation of motion for $\Sigma$ given in Eq.~(\ref{eq:eom_Sigma}) 
with the condition $\Sigma = \Sigma^\dagger$.
Note also that $\Sigma$ does not depend on $t$ and $x^3$ because neither does $H_\star$.
In summary, for a given solution $H$ and $W_m$, this equation determines 
$\Sigma(x^1,x^2) = \mp W_0(x^1,x^2)$.

A final comment is on the last line of the second equality of Eq.~(\ref{eq:Bogo}). Since $\Sigma = \Sigma^\dagger$,
it can be written as
\beq
\pm \Sigma \left\{-\frac{2}{g^2}\D_mF_{0m} - i \left(H(\D_0 H)^\dagger - \D_0H H^\dagger\right)\right\}.
\eeq
This is precisely zero due to the Gauss law (\ref{eq:Gauss}), so that there are no contributions to
the energy density.

\subsection{A low energy effective theory in the massless case}
\label{sec:low energy}

There is a clear-cut view point for understanding the BPS DNALVs~\cite{Collie:2008za}. It is 
a low energy effective theory on the vortex world volume. 
As explained in Sec.~\ref{sec:bps NA vortex}, the single NALVS has the massless
moduli $\mathbb{C}P^{\NC-1}$ in the case of $M = 0$. 
In this case, the low energy effective theory is  
$1+1$ dimensional non-linear sigma model whose target space is $\mathbb{C}P^{\NC-1}$.

We will study the massless case $M=0$ in this subsection.
Let us see the simplest case of $\NC=2$ as a concrete example.
First of all, 
we have to specify the orientational moduli in the solution.
It turns out that the singular gauge is simpler. 
So we transform $H$ and $W_{1,2}$ given in Eqs.~(\ref{eq:NA_BPS_ansatz_1}) 
and (\ref{eq:NA_BPS_ansatz_2}) to the following ones
by the gauge transformation $H \to V H$ and $W_\mu \to VW_\mu V^\dagger + i (\p_\mu V) V^\dagger$
with $V = {\rm diag.}(e^{-i\theta},1)$:
\beq
H = v \left(
\begin{array}{cc}
h(r) & 0\\
0 & 1
\end{array}
\right),\qquad
W_1 + i W_2 = \left(
\begin{array}{cc}
-i\frac{w(r)-1}{r}e^{i\theta} & 0\\
0 & 0
\end{array}
\right).
\label{eq:HW0}
\eeq
In order to specify the orientational moduli, we transform these by an $SU(2)_{\rm C+F}$ transformation as
\beq
H = U \left[
v\left(
\begin{array}{cc}
h  & 0 \\
0 & 1
\end{array}
\right) \right]
U^{-1},
\qquad
W_1+iW_2 = U
\left(
\begin{array}{cc}
-i\frac{w-1}{r}e^{i\theta} & 0\\
0 & 0
\end{array}
\right)
U^{-1}.
\label{eq:HW}
\eeq
Here $U$ is an $SU(2)_{\rm C+F}$ matrix which we parametrize by a complex parameter $\phi\in\mathbb{C}$
as
\beq
U = \left(
\begin{array}{cc}
\frac{1}{\sqrt{1+|\phi|^2}} & \frac{\phi}{\sqrt{1+|\phi|^2}}\\
-\frac{\phi^*}{\sqrt{1+|\phi|^2}} & \frac{1}{\sqrt{1+|\phi|^2}}
\end{array}
\right) \in SU(2)_{\rm C+F}.
\label{eq:ori_H}
\eeq
Note that the phase of the left-upper component of $U$ can be always fixed to be real and positive by using
$\exp(i\alpha\sigma_3/2) \in U(1)_{\rm C+F}$ transformation: $U \to U\exp(i\alpha\sigma_3/2)$. 
Note that this $U(1)_{\rm C+F}$ does not change $H$ and $W_{1,2}$
given in Eq.~(\ref{eq:HW0}). In other words, this is an isotropy of the moduli space.
Hence, the moduli space is isomorphic to $SU(2)_{\rm C+F}/U(1)_{\rm C+F} \simeq \mathbb{C}P^1$ and 
$\phi$ is an inhomogeneous coordinate of $\mathbb{C}P^1$. 

In order to derive a low energy
effective theory for the orientational moduli in the massless case $M=0$, we promote the complex parameter
$\phi$ to be a field $\phi(x^0,x^3)$ on the vortex world-volume~\cite{Gibbons:1986df},
which gives us $x^0$ and $x^3$ dependence in $H$ and $W_{1,2}$. 
We also need to clarify $x^0,x^3$ dependence of  $W_{0,3}$ while we set $\Sigma$ to zero. 
To this end, we make an Ansatz as follows~\cite{Shifman:2004dr}
\beq
W_\alpha(x^0,x^3) = - 2i \lambda(r) \left(U\frac{\sigma_3}{2}U^{-1}\right) \p_\alpha\left(U\frac{\sigma_3}{2}U^{-1} \right),\qquad (\alpha = 0,3),
\label{eq:W0_eff}
\eeq
where $x^{0,3}$ dependence enters through $U$, and $\lambda(r)$ is an unknown function of $r$.
Finally, 
plugging $H(x^1,x^2;\phi(x^3,t))$, $W_{1,2}(x^1,x^2;\phi(x^3,t))$
and $W_{0,3}(x^1,x^2;\phi(x^3,t))$  into the Lagrangian (\ref{eq:lag_U(N)}), and
integrating the quadratic derivative terms in $x^3$ and $t$ over $x^1$ and $x^2$, we get 
\beq
{\cal L}_{\rm eff}^{(2)} = \int dx^1dx^2\ \Tr\left[
-\frac{1}{g^2}F_{m\alpha}F^{m\alpha} + \D_\alpha H (\D^\alpha H)^\dagger
\right] 
= \beta \frac{\p_\alpha \phi \p^\alpha \phi^*}{(1+|\phi|^2)^2},
\label{eq:cp1}
\eeq 
with $m=1,2$, $\alpha = 0,3$ and
\beq
\beta = \frac{4\pi}{g^2} \int d\rho\ \rho
\left[ 
\lambda '^2+(1-\lambda)^2 \frac{(1-w)^2}{\rho^2}
+
\frac{\lambda^2}{2} \left(h^2+1\right)+(1-\lambda) (1-h)^2
\right].
\label{eq:beta}
\eeq
We regard this as a potential for $\lambda(\rho)$ which is minimized by
\beq
\lambda'' + \frac{\lambda'}{\rho} +\frac{(1-\lambda) (1-w)^2}{\rho^2}
+\frac{1}{2}  \left((1+h^2)(1-\lambda)-2h \right) = 0.
\eeq
Note that the same equation can also be derived by plugging
$H$ and $W_\mu$ given above into the Gauss law (\ref{eq:Gauss}) with $\Sigma = 0$.
By using the BPS equation (\ref{eq:NA_BPS_axial}), this can be solved by~\cite{Shifman:2004dr}
\beq
\lambda(\rho) = 1-h(\rho),
\label{eq:sol_SY}
\eeq
where $h(\rho)$ is the background profile function shown in Fig.~\ref{fig:BPS_NALVS}.
Plugging this back into Eq.~(\ref{eq:beta}), and using the BPS equations (\ref{eq:NA_BPS_axial}), 
we get an analytic result~\cite{Shifman:2004dr}
\beq
\beta = \frac{4\pi}{g^2}\int d\rho\ \frac{d(w-1)}{d\rho} = \frac{4\pi}{g^2}.
\eeq

In the expression (\ref{eq:cp1}), the $U(1)$ global symmetry 
$\phi \to e^{i\eta}\phi$,
which is a subgroup
of isometry of $\mathbb{C}P^1 \simeq S^2$ is manifest. The corresponding Noether current is given by
\beq
j^3_\alpha = i \beta \frac{\phi\p_\alpha\phi^* - \phi^*\p_\alpha \phi}{(1+|\phi|^2)^2}.
\eeq
Thus the corresponding Noether charge per unit length is given by
\beq
q^3 = i\beta \frac{\phi\dot\phi^* - \phi^*\dot\phi}{(1+|\phi|^2)^2}.
\eeq
Note that an origin of this $e^{i\eta} \in U(1)$ symmetry in the effective theory can be traced to
$\exp(-i\eta\sigma_3/2) \in SU(2)_{\rm C+F}$ symmetry of the original theory in $3+1$ dimensions.
This can be seen by transforming $U$ given in Eq.~(\ref{eq:ori_H}) with the $U(1)_{\rm F}$
transformation associated with the $U(1)_{\rm C}$ as
\beq
U  \to e^{i\eta\frac{\sigma_3}{2}}
U e^{-i\eta\frac{\sigma_3}{2}}
\quad \Leftrightarrow \quad
\phi  \to e^{i\eta}\phi.
\label{eq:u1s}
\eeq
Therefore, $q^3$ should be identified with
$Q^3/2$
defined in Eq.~(\ref{eq:Q_SU(2)}). Indeed, one can directly check this by calculating 
$Q^3$ from Eqs.~(\ref{eq:Q_SU(2)}),
(\ref{eq:HW}),  (\ref{eq:W0_eff}), and (\ref{eq:sol_SY}) as
\beq
Q^3 
= \frac{i  \left(\phi^* \dot\phi-\phi \dot\phi^*\right)}{(1+|\phi|^2)^2} 
\int dx^1dx^2\  2v^2 \left(2 h \lambda+(h-1)^2\right)
= 2 q^3,
\label{eq:Q=2q}
\eeq
where we have also used Eq.~(\ref{eq:NA_BPS_axial}).

\subsection{A low energy effective theory in the massive case}

Next, we turn on non-zero mass matrix 
\beq
M = \frac{m}{2}\sigma_3,\qquad (m > 0).
\eeq
Then, we cannot set $\Sigma$ to zero anymore.
The mass term explicitly breaks the flavor
symmetry $SU(2)_{\rm F}$ down to $U(1)_{\rm F}$ subgroup. However,
as long as the masses are kept sufficiently small ($m \ll \mu$), 
$SU(2)_{\rm F}$ can be dealt with as an approximate symmetry,
and the low energy effective theory remains $\mathbb{C}P^1$ non-linear sigma model.
Instead, the small symmetry breaking term generates a small effective potential. Namely the effective
theory becomes {\it massive} $\mathbb{C}P^1$ sigma model.
In order to derive the effective potential, we make the following Ansatz for $\Sigma$~\cite{Shifman:2004dr}
\beq
\Sigma(t,x^3) = \frac{m}{2}\left\{(1-\sigma(r))\sigma_3 
+ 4\sigma(r) \Tr\left[\left(U\frac{\sigma_3}{2}U^{-1}\right)\frac{\sigma_3}{2}\right]\left(U\frac{\sigma_3}{2}U^{-1}\right)
\right\},
\eeq
with an unknown function $\sigma(r)$. We plug this into the Lagrangian  (\ref{eq:lag_U(N)}). Then
the quadratic term in $m$ yields the desired effective potential
\beq
{\cal V}_{\rm eff} &=& - \int dx^1dx^2\ \Tr \left[
\frac{1}{g^2}\left(\D_1\Sigma(\D^1\Sigma)^\dagger + \D_2\Sigma(\D^2\Sigma)^\dagger \right)
- \left(\Sigma H - H M\right) \left(\Sigma H - H M\right)^\dagger\right] \non
&=& \chi \frac{m^2 |\phi|^2}{(1+|\phi|^2)^2}.
\label{eq:eff_pot}
\eeq 
Here the coefficient $\chi$ is given by the following integral
\beq
\chi = \frac{4\pi}{g^2} \int d\rho\ \rho \left[
\sigma'^2+ (1-\sigma)^2\frac{(1-w)^2}{\rho^2}
+ 
\frac{\sigma^2}{2}\left(h^2+1\right) +(1-\sigma) (1-h)^2
\right].
\eeq
Note that this is exactly the same form as $\beta$ given in Eq.~(\ref{eq:beta}), so that
the coefficient $\chi$ is also minimized by $\sigma(r) = \lambda(r) = 1- h(r)$~\cite{Shifman:2004dr}. 
Hence, we find
\beq
\chi = \beta = \frac{4\pi}{g^2}.
\eeq
In summary, the effective Lagrangian is sum of the kinetic term given in Eq.~(\ref{eq:cp1}) and 
the potential term given in Eq.~(\ref{eq:eff_pot})
\beq
{\cal L}_{\rm eff}^{(2)} = \beta \left[ \frac{\p_\alpha \phi \p^\alpha \phi^*}{(1+|\phi|^2)^2} 
-  \frac{m^2 |\phi|^2}{(1+|\phi|^2)^2}\right].
\eeq
There are two vacua: the one for $\phi = 0$ and the other for $\phi=\infty$. The former corresponds to
$U={\bf 1}_2$ given in Eq.~(\ref{eq:ori_H}), so it gives NALVS living in the left-upper corner,
see Eq.~(\ref{eq:HW}). On the other hand, $\phi=\infty$ corresponds to $U=i\sigma_2$, so it gives
NALVS living in the right-bottom corner. We identify the former and  the latter to the north and the south
poles of $\mathbb{C}P^1$, respectively. All the other configurations are lifted by the non-zero
mass $M$.

It is also useful to rewrite everything in terms of spherical coordinates
\beq
\phi = e^{i\Phi}\tan\frac{\Theta}{2},\qquad 0 \le \Theta \le \pi,\quad 0 \le \Phi < 2\pi.
\label{eq:ch_coordinate}
\eeq
Then the massive $\mathbb{C}P^1$ non-linear sigma model is expressed by
\beq
{\cal L}_{\rm eff}^{(2)} = \frac{\beta}{4} \left[ \partial_\alpha \Theta \partial^\alpha \Theta 
+ \partial_\alpha\Phi\partial^\alpha \Phi \sin^2\Theta - m^2 \sin^2\Theta\right].
\eeq
In terms of the spherical coordinate, the two vacua are $\Theta=0$ and $\Theta = \pi$.

\subsection{The BPS DNALVS from the effective field theory}
\label{sec:dnalvs_eff_th}

We are now ready to reconstruct the BPS DNALVS in the effective theory.
As can be seen from Eq.~(\ref{eq:H_dyonic}), DNALVS can be generated by time-dependent
flavor rotation $e^{imt\sigma_3/2} \in U(1)_{\rm F}$. Combining this and Eq.~(\ref{eq:u1s}),
we see that $\phi$ should be transformed as $\phi \to \phi e^{imt}$. Namely, we should 
have a time dependence of $\Phi$ as $\Phi(t) = mt$.
Actually, the BPS nature of the dyonic configuration tells us that this is the case
\beq
H_{\rm eff}^{(2)} 
&=& \frac{\beta}{4} \left[
(\p_0 \Theta)^2 + (\p_3\Theta)^2 +\left\{ (\p_0\Phi)^2 + (\p_3\Phi)^2\right\}\sin^2\Theta
+ m^2 \sin^2\Theta
\right] \non
&=& \frac{\beta}{4}\left[
(\p_0 \Theta)^2 + (\p_3\Theta)^2 +(\p_3\Phi)^2\sin^2\Theta
+ \left(\p_0\Phi \mp  m\right)^2 \sin^2\Theta
+ 2m\p_0\Phi \sin^2\Theta
\right]\non
&\ge& m |q^3| ,
\eeq
with the conserved Noether charge per unit length in the spherical coordinate 
\beq
q^3 = \frac{\beta}{2} \p_0\Phi \sin^2\Theta.
\label{eq:angular_momentum}
\eeq
The Bogomol'nyi bound is saturated when
\beq
\p_0\Theta = \p_3\Theta = \p_3\Phi=0,\quad \p_0\Phi = \pm m,
\eeq
which gives the desired time dependence
\beq
\Phi = \pm mt.
\label{eq:Phi_BPS}
\eeq
Thus, the BPS tension of this state in the low energy effective theory reads $m|q^3|$. 
The non-dyonic NALVS gains this as an additional
tension, so that the total tension of the single DNALVS becomes
\beq
T^{\rm (BPS)} = 2\pi v^2 + m |q^3|.
\label{eq:BPS_dv_mass_eff}
\eeq
Let us compare this with the BPS tension formula (\ref{eq:BPS_dv_mass}) in the original theory.
Since we have $M = m\sigma_3/2$, Eq.~(\ref{eq:BPS_dv_mass}) with $k=1$ gives
$T = 2\pi v^2 + m|Q^3|/2$. Since $|Q^3| =2 |q^3|$ from  Eq.~(\ref{eq:Q=2q}),  we see 
that Eqs.~(\ref{eq:BPS_dv_mass}) and (\ref{eq:BPS_dv_mass_eff}) are indeed identical.

Assuming that $\Theta$ and $\Phi$ are function of $t$ only,
we can eliminate $\Phi$ from the effective Lagrangian by using the equation of motion,
\beq
{\cal L}_{\rm eff}^{(2)} = \frac{\beta}{4}\left[
\p_0\Theta^2 - m^2 \sin^2\Theta - \frac{4(q^3)^2}{\beta^2\sin^2\Theta}
\right].
\eeq
There are two minima of the effective potential 
${\cal V}_{\rm eff} = m^2 \sin^2\Theta + 4(q^3)^2 / \beta^2\sin^2\Theta$ at
\beq
\sin^2 \Theta_0 = \frac{2q^3}{m\beta}\quad \to \quad 
\Theta_0 = \pm \sin^{-1}\sqrt{\frac{2q^3}{m\beta}}.
\eeq
Thus, for a given $q^3$, there exist two BPS dyonic solutions~\cite{Collie:2008za}.

\begin{figure}[ht]
\begin{center}
\includegraphics[height=7cm]{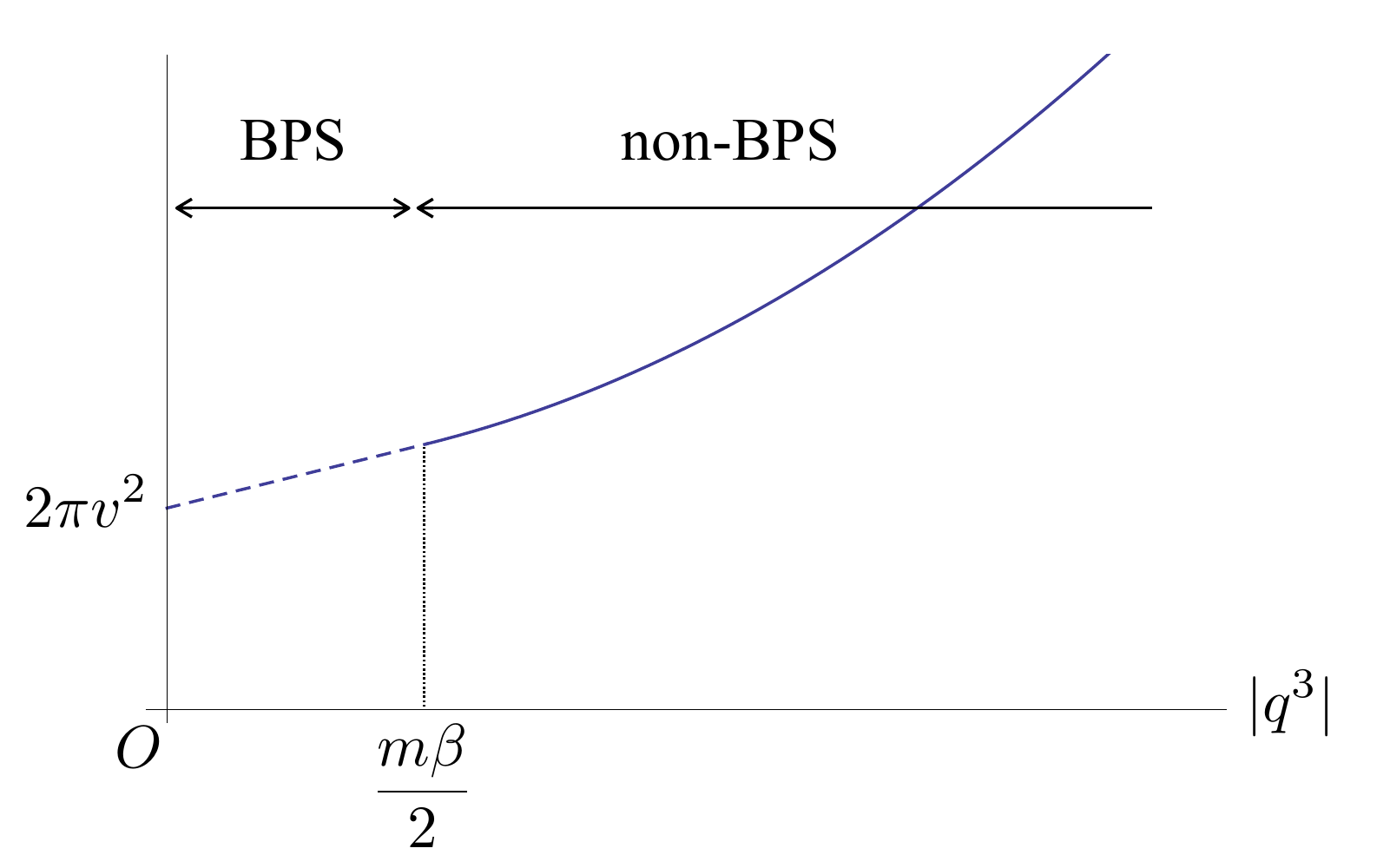}
\caption{Tension of the dyonic non-Abelian vortex as function of the angular momentum $q^3$.
The BPS solutions with $|q^3| \le m\beta/2$ is doubly degenerate while the non-BPS solutions with
$|q^3| > m\beta/2$ are not degenerate.}
\label{fig:mass_formula}
\end{center}
\end{figure}
Below Eq.~(\ref{eq:BPS_dv_mass}), we mentioned 
that the BPS DNALVS cannot have an arbitrary large Noether charge.
Now we are ready to explain this as follows~\cite{Collie:2008za}.
The Noether charge for BPS state is given in Eq.~(\ref{eq:angular_momentum}). Thus, we have
\beq
\left|q^3\right|_{\rm BPS} = \left|\frac{\beta}{2}\p_0\Phi\sin^2\Theta\right|_{\rm BPS}
 \le \left|\frac{\p_0\Phi \beta}{2}\right|_{\rm BPS} = \frac{m\beta}{2} \equiv q_{\rm c}^3.
\eeq
Therefore, any configurations with the conserved charge greater than the critical charge $q_{\rm c}^3$
cannot be BPS state. For such solutions, a centrifugal force in
the internal space forces $\Theta$ to be $\pi/2$.
Namely, while there are two degenerate configurations for BPS states with $|q^3| \le q_{\rm c}^3$,
there is a unique non-BPS state for $|q^3| > q_{\rm c}^3$.
From Eq.~(\ref{eq:angular_momentum}) with $\Theta = \pi/2$, 
the non-BPS solution is given by
\beq
\Phi^{\text{non-BPS}} = \omega t,\qquad q^3 = \frac{\beta\omega}{2}. 
\label{eq:Phi_nonBPS}
\eeq
Thus, gain in the tension of the solution above $q_{\rm c}$ is
\beq
\delta T = 
\frac{\beta}{4}\left[ \left(\frac{2q^3}{\beta}\right)^2 + m^2\right] = \frac{(q^3)^2}{\beta} + \frac{\beta m^2}{4},
\eeq
and the total tension of the non-BPS DNALVS reads
\beq
T^{\text{(non-BPS)}} = 2\pi v^2 + \frac{(q^3)^2}{\beta} + \frac{\beta m^2}{4}.
\label{eq:tension_nBPS}
\eeq
In summary, DNALVS continuously changes from BPS to non-BPS at $|q^3| = m\beta/2$,
see Fig.~\ref{fig:mass_formula}.

While the BPS tension formula (\ref{eq:BPS_dv_mass_eff}) is of the dyonic-instanton type, 
Eq.~(\ref{eq:tension_nBPS}) for the
non-BPS state is neither of the dyon nor of the dyonic-instanton type. 
However,
note that the results obtained in this section are valid only for small $\omega \lesssim m \ll \mu$ since they are all
derived from the low energy effective action.
In Sec.~\ref{sec:non_BPS_dyonic_U2_vortex}, we will focus on the non-BPS states above $|q^3| = m\beta/2$,
especially on the case with $m=0$ where all the dyonic solutions are non-BPS.

\subsection{Spiral dyonic solitons}
\label{sec:lorentz boost}

\begin{figure}[t]
\begin{center}
\includegraphics[width=17cm]{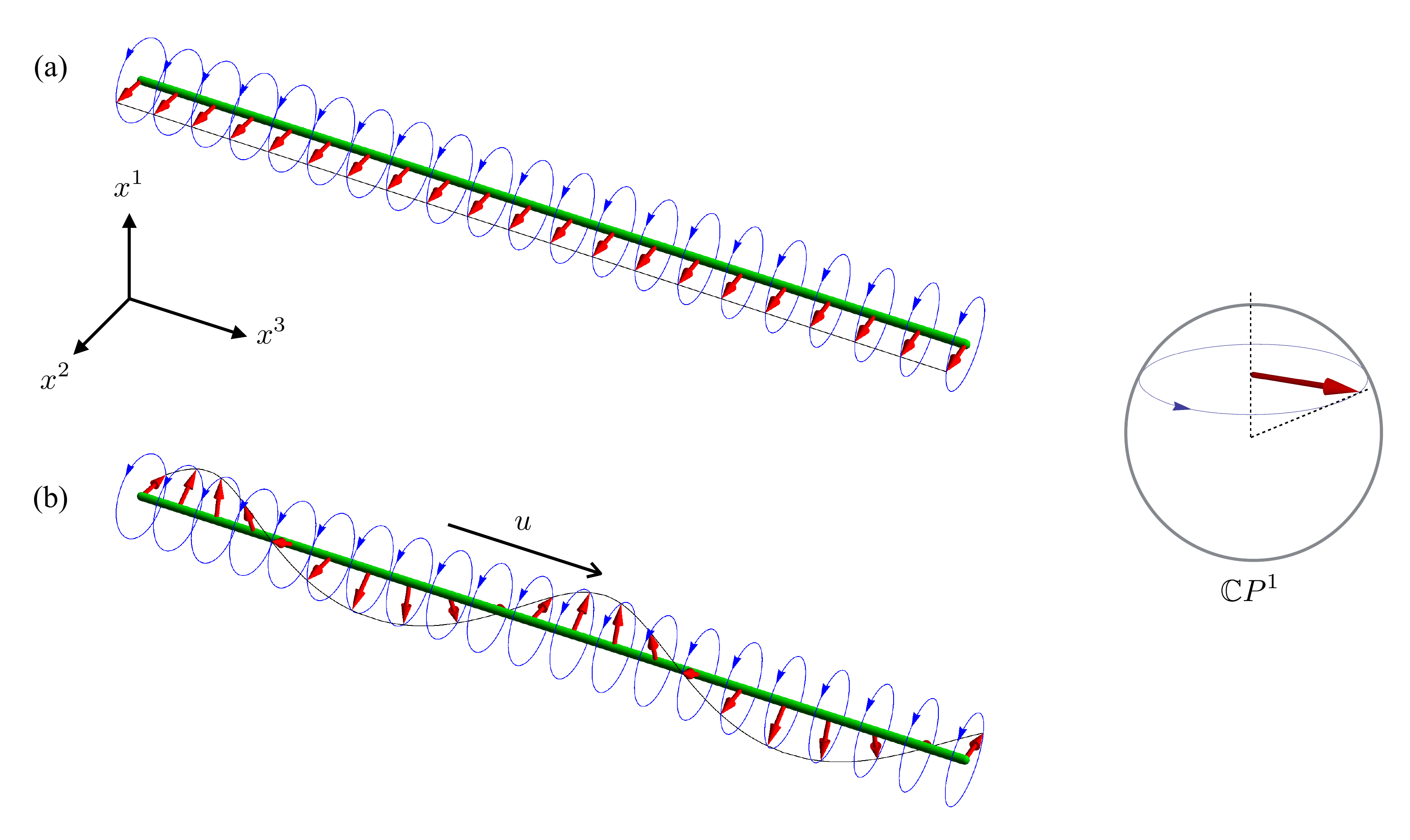}
\caption{(a) The trivial DNALVS and (b) the spiral DNAVLS.
The green thick line stands for the non-Abelian vortex string extending along the $x^3$ axis in the real space 
while the red arrows show
the internal orientation $\Phi$ at each $x^3$ of the strings.}
\label{fig:spiral}
\end{center}
\end{figure}

DNALVS considered so far has no $x^3$ dependence, and 
it trivially extends along the $x^3$ axis.
Therefore, all the results in the previous sections also hold in $2+1$ dimensions without any changes.
Now let us generate new solutions which depend on $x^3$. Namely, we are going to study
genuine $3+1$ dimensional configurations.  An easy way for finding such solutions
is boosting the $x^3$-independent solution along the $x^3$ direction.
We begin with the $x^3$-independent solution, namely DNALVS; 
$H(t,x^1,x^2) = H_\star(x^1,x^2)e^{\pm i Mt}$ given  in Eq.~(\ref{eq:H_dyonic}),
the solution $W_{1,2}(x^1,x^2)$  of
Eq.~(\ref{eq:NA_vor_BPS_eq_again}) which is independent of both $t$ and $x^3$ ,
$\Sigma(x^1,x^2) = \mp W_0(x^1,x^2)$ are also $t$- and $x^3$-independent which are determined by
Eq.~(\ref{eq:eq_Sigma}), and $W_3 = 0$. 
Now we boost these configurations. This can be done by just replacing $t$ by  $(t-ux^3)/\sqrt{1-u^2}$ with
$0 \le u < 1$.
Thus, the boosted solution is given by
\beq
H &=& H_\star(x^1,x^2) \exp\left(iM\frac{t - u x^3}{\sqrt{1-u^2}}\right),
\label{eq:boost_phase}\\
W_{1,2} &=& W_{1,2}(x^1,x^2),\\
W_0 &=& \mp \frac{1}{\sqrt{1-u^2}} \Sigma(x^1,x^2),\\
W_3 &=& \mp \frac{-u}{\sqrt{1-u^2}} \Sigma(x^1,x^2),
\label{eq:boost1}
\eeq
where $W_{1,2}(x^1,x^2)$ and $\Sigma(x^1,x^2)$ in the right hand side are the solutions
of Eqs.~(\ref{eq:NA_vor_BPS_eq_again}) and (\ref{eq:eq_Sigma}).
The Noether charge per unit length is transformed  as
\beq
Q^3 \to 
\frac{1}{\sqrt{1-u^2}}\, Q^3.
\eeq
Plugging this into the BPS tension formula (\ref{eq:BPS_dv_mass}), we get the following
formula for the boosted vortex string
\beq
T = 2\pi v^2 k + \frac{1}{\sqrt{1-u^2}}\,\left|\Tr[MQ]\right|.
\label{eq:BPS_sdv_mass}
\eeq
Note that only the second term is transformed while the first term remains unchanged because
it corresponds to the vortex tension, which is nothing to do with
$t$ and $x^3$. In order to understand this configuration better, let us look at $t$- and 
$x^3$-dependence in the phase of $H$, see Eq.~(\ref{eq:boost_phase}).
The phase can be identified with $t$- and $x^3$-dependent flavor transformation, so that
$\theta = \frac{t - u x^3}{\sqrt{1-u^2}}$ is nothing but rotating angle on $\sigma^3$-axis in the 
internal orientation moduli space.  Therefore, if $u=0$, the internal orientation uniformly rotates
with an angular velocity $m$, see Fig.~\ref{fig:spiral} (a). 
On the other hand, once $u \neq 0$, the rotating angle depends on
$x^3$, so the orientation spirals along $x^3$-axis with a phase velocity $u$, see Fig.~\ref{fig:spiral} (b).

The same can be seen in the low energy effective theory on the vortex string world volume.
The BPS solution corresponding to DNALVS is given by
$\Phi = mt$ and $\Theta = \Theta_0$. This is transformed by a Lorentz boost as
\beq
\Phi(t,x^3) = m \frac{t-ux^3}{\sqrt{1-u^2}}, \quad \Theta = \Theta_0.
\label{eq:boost2}
\eeq
Of course, this solves the equations of motion
\beq
\p_\alpha\p^\alpha \Theta - \left(\p_\alpha\Phi\p^\alpha\Phi - m^2\right)\sin 2\Theta = 0,\qquad
\p_\alpha\left(\p^\alpha \Phi\sin^2\Theta\right) = 0.
\eeq
The conserved charge is given by
\beq
q^3 = \frac{\beta}{2}\p_0\Phi \sin^2\Theta = \frac{1}{\sqrt{1-u^2}}\,\frac{m\beta}{2}\sin^2\Theta_0,
\eeq
and an increment in the tension reads
\beq
\delta T = \frac{\beta}{4} \left((\p_0\Phi)^2 + (\p_3\Phi)^2 + m^2\right)\sin^2\Theta_0 = 
\frac{1}{\sqrt{1-u^2}}\, m|q^3|.
\eeq
This is consistent with the above result in Eq.~(\ref{eq:BPS_sdv_mass}).

A comment on the spinning non-Abelian vortex strings:
In Ref.~\cite{Brihaye:2007pd}, a sort of non-Abelian vortex strings whose profile functions depend not
only on $x^1,x^2$ but also $t,x^3$ were found in the $U(\NC)$ Yang-Mills-Higgs model with
$\NF$ flavors. This is a kind of spinning soliton and such string exists only in the case $\NF > \NC$.
Namely, it is the spinning non-Abelian {\it semi-local} vortex strings.
The spinning non-Abelian semi-local vortex strings is always non-BPS, and
the internal orientation does not rotate in neither $t$ nor $x^3$
unlike the spiral DNALVS studied in this subsection.

\section{Non-BPS dyonic non-Abelian local vortex strings}
\label{sec:non_BPS_dyonic_U2_vortex}

In the previous section, we have seen that the dyonic extension makes the BPS NALVS be non-BPS, 
if the Noether charge exceeds the critical value $|Q^3| = m\beta = 4\pi m/g^2$. This fact indicates, if $m=0$, that any DNALVSs are non-BPS states. 
We will investigate the non-BPS DNALVSs in the massless case $M=0$ in this section,
which have not been studied in the literature.

\subsection{Non-BPS solutions}

Let us first see what happens if we set $M=0$ in the Bogomol'nyi completion (\ref{eq:Bogo}).
Since the Noether charge cannot contribute to the tension, we just return to the tension formula 
for the non-dyonic BPS NALVS
\beq
H \ge \int dx^1dx^2\ \Tr\left[-v^2 F_{12} \right].
\eeq
The BPS equations 
(\ref{eq:NA_vor_BPS_eq_again}) -- (\ref{eq:NA_vor_BPS_eq_new}) with $M=0$ allow only
$x^0$- and $x^3$-independent configurations. Namely, they are nothing but 
the non-dyonic BPS NALVSs explained in Sec.~\ref{sec:bps NA vortex}.
This indicates that any time-dependent vortex strings in the massless case 
are non-BPS states, which is consistent with
the previous observation from the view point of $1+1$ dimensional effective theory 
in Sec.~\ref{sec:dnalvs_eff_th}.

Therefore, instead of solving the BPS equations, we have to solve the full equations of motion given
in Eqs.~(\ref{eq:eom_W}) -- (\ref{eq:eom_H}) with $M=0$ for 
time-dependent configurations.
Since we are interested in $x^3$-independent configurations, we will set $W_3=0$ and $\p_3=0$ hereafter.
Furthermore, we impose $\Sigma = 0$ which solves Eq.~(\ref{eq:eom_Sigma}) for $M=0$.
Hence, we are left with the following equations 
\beq
\D_\mu\D^\mu H + \frac{g^2}{2}(HH^\dagger - v^2{\bf 1}_2)H = 0,
\label{eq:eom1}\\
\frac{1}{g^2}\D_\mu F^{\mu \nu}
+ i \left(H (\D^\nu H)^\dagger - (\D^\nu H)H^\dagger\right)= 0.
\label{eq:eom2}
\eeq

In order to solve these differential equations, we need to make an appropriate Ansatz for
$H$ and $W_{1,2,0}$.  To this end, we first note that the flavor symmetry $SU(2)_{\rm F}$ is manifest
because we have $M=0$. Hence, we have three conserved flavor charges $Q^a$ ($a=1,2,3$) associated
with the generators of $SU(2)_{\rm F}$. Without loss of generality, we will consider configurations
with $Q^a = Q\delta^{3a}$. 
Since the non-Abelian vortex has the orientational moduli $\mathbb{C}P^1 \simeq S^2$ for
the case of $\NC=2$, it may be instructive to imagine a free particle confined on a sphere.
A free particle on a sphere always moves along a great circle, and the motion is specified by
angular momentum.
In particular, a particle with $Q^a = Q\delta^{a3}$ rotates on the equator. 
In terms of the spherical coordinate defined in Eq.~(\ref{eq:ch_coordinate}),
the motion is expressed as
$\Theta = \pi/2$ and  $\Phi = \omega t$.
With these observation in mind, let us now make an appropriate Ansatz.
We begin with the BPS non-dyonic NALVS given in Eqs.~(\ref{eq:NA_BPS_ansatz_1})
and (\ref{eq:NA_BPS_ansatz_2}) for $\NC=2$. Since the configuration is diagonal, any
flavor transformation generated by $T^3$ are absorbed by $U(2)_{\rm C}$ gauge transformation.
In other words, it corresponds to a particle on the north pole of $S^2$.
Therefore, we first need to transform it
by $\pi/2$ on $T^2=\sigma^2/2$, by which we send a particle from the north pole to a point  on
the equator. Now we are ready to make an Ansatz as~\cite{Eto:2012qda}
\beq
H(x^1,x^2,t) &=& \tilde U^\dagger(t) \left[ v\left(
\begin{array}{cc}
h_1(r)e^{i\theta} & 0 \\
0 & h_2(r)
\end{array}
\right)\right]
\tilde U(t),\\
W_1(x^1,x^2,t) + i W_2(x^1,x^2,t) &=& \tilde U^\dagger(t)
\left[
- \frac{i e^{i\theta}}{r} \left(
\begin{array}{cc}
w_1(r) & 0\\
0 & w_2(r)
\end{array}
\right)
\right]
\tilde U(t),
\eeq
with
\beq
\tilde U(t) =  \exp\left(i\frac{\pi}{2}T_2\right)\exp\left(i\omega t T_3\right)
= \frac{1}{\sqrt{2}}\left(
\begin{array}{cc}
 e^{\frac{i\omega t}{2}} & -  e^{-\frac{i\omega t}{2}} \\
  e^{\frac{i\omega t}{2}} &  e^{-\frac{i\omega t}{2}} \\
\end{array}
\right).
\label{eq:Ut}
\eeq
Remember that the right-bottom elements of $H$ and $W_{1,2}$ are trivial ($h_2=1,\ w_2 = 0$) 
for the BPS non-dyonic NALVS with
$\omega = 0$. On the contrary, for time-dependent dyonic configurations, 
it will turn out that they are non-trivial, so we leave
the right-bottom elements to be unknown profile functions.
We also need to make an Ansatz for $W_0$~\cite{Eto:2012qda}
\beq
W_0(x^1,x^2,t) &=& \tilde U^\dagger(t)
\left[
\frac{\omega}{2}
\left(
\begin{array}{cc}
0 & 1-e^{i\theta}f(r) \\
1-e^{-i\theta}f(r) & 0
\end{array}
\right)
\right]\tilde U(t).
\eeq
Plugging $H$ and $W_{0,1,2}$ into the equations of motion
(\ref{eq:eom1}) and (\ref{eq:eom2}), we have  the following ordinary differential equations
\beq
{\rm eq}_1 \!\!&\equiv&\!\! h_1''  + \frac{h_1'}{\rho} 
-\left(\frac{h_1^2-1}{2}  +\frac{(1-w_1)^2}{\rho^2}\right)h_1
+ \frac{\tilde \omega^2}{2}\left[
\frac{(1-f)^2}{2} h_1
+ (h_1-h_2)f\right] = 0,
\label{eq:eom1}\\
{\rm eq}_2 \!\!&\equiv&\!\! 
h_2''  + \frac{h_2'}{\rho} 
-\left(\frac{h_2^2-1}{2} +\frac{w_2^2}{\rho^2}\right)h_2
+ \frac{\tilde \omega^2}{2}\left[
\frac{(1-f)^2}{2} h_2
- (h_1-h_2)f\right] = 0,
\label{eq:eom2}\\
{\rm eq}_3 \!\!&\equiv&\!\! 
w_1'' - \frac{w_1'}{\rho} + h_1^2 (1-w_1) - \frac{\tilde \omega^2}{2} f^2 (1-w_1+w_2) = 0,
\label{eq:eom3}\\
{\rm eq}_4 \!\!&\equiv&\!\! 
w_2'' - \frac{w_2'}{\rho} - h_2^2 w_2 + \frac{\tilde \omega^2}{2} f^2 (1-w_1+w_2) = 0,
\label{eq:eom4}\\
{\rm eq}_5 \!\!&\equiv&\!\! 
f'' +\frac{f'}{\rho} - \frac{(1-w_1+w_2)^2}{\rho^2} f -
 \left(\frac{f}{2}  (h_1-h_2)^2 - (1-f) h_1 h_2\right) = 0,
 \label{eq:eom5}
\eeq
with a dimensionless parameter
\beq
\tilde \omega = \frac{\omega}{\mu}.
\eeq
As we will see below, an absolute value of $\tilde \omega$ should be less than one.
We solve these differential equations for $h_{1,2}(\rho)$, $w_{1,2}(\rho)$ and $f(\rho)$ 
with the following boundary conditions
\beq
h_1(0) = 0,\quad h_2'(0) = 0,\quad w_1(0)=0,\quad w_2(0) = 0,\quad f(0)=0,
\label{eq:bc0}
\eeq
and
\beq
h_1(\infty) = 1,\quad h_2(\infty) = 1,\quad w_1(\infty)=1,\quad w_2(\infty) =0,\quad f(\infty) = 1.
\label{eq:bc_infty}
\eeq
The energy density is expressed as
\beq
{\cal H} = g^2v^4\left({\cal K}_{\rm s} + {\cal K}_{\rm t} + {\cal V}\right),
\eeq
with
\beq
{\cal K}_{\rm s} &=& \frac{1}{g^2v^4}
 {\rm Tr}
\left[
\frac{1}{g^2}F_{12}^2 + \D_m H(\D_mH)^\dagger
\right] \non
&=& \frac{w_1'{}^2+w_2'{}^2}{\rho^2}
+ \left(h_1'{}^2+h_2'{}^2\right)
+\frac{h_1^2 (1-w_1)^2+h_2^2 w_2^2}{\rho^2},\\
{\cal K}_{\rm t} &=& \frac{1}{g^2v^4}
{\rm Tr}\left[\frac{1}{g^2}F_{0m}^2
+ \D_0H (\D_0H)^\dagger\right] \non
&=& \frac{\tilde\omega^2}{2}\bigg[
f'^2 + \frac{f^2 (-w_1+w_2+1)^2}{\rho^2}
+ \frac{1}{2}\left(
(1+f^2)(h_1^2+h_2^2) -4fh_1h_2 \right)
\bigg],\\
{\cal V} &=& \frac{1}{g^2v^4} \Tr\left[
\frac{g^2}{4}\left(HH^\dagger - v^2{\bf 1}_2\right)^2\right]\non
&=& \frac{1}{4} \left(\left(h_1^2-1\right)^2+\left(h_2^2-1\right)^2\right),
\eeq
where the prime stands for a derivative in terms of $\rho$.
\begin{figure}[t]
\begin{center}
\includegraphics[width=13cm]{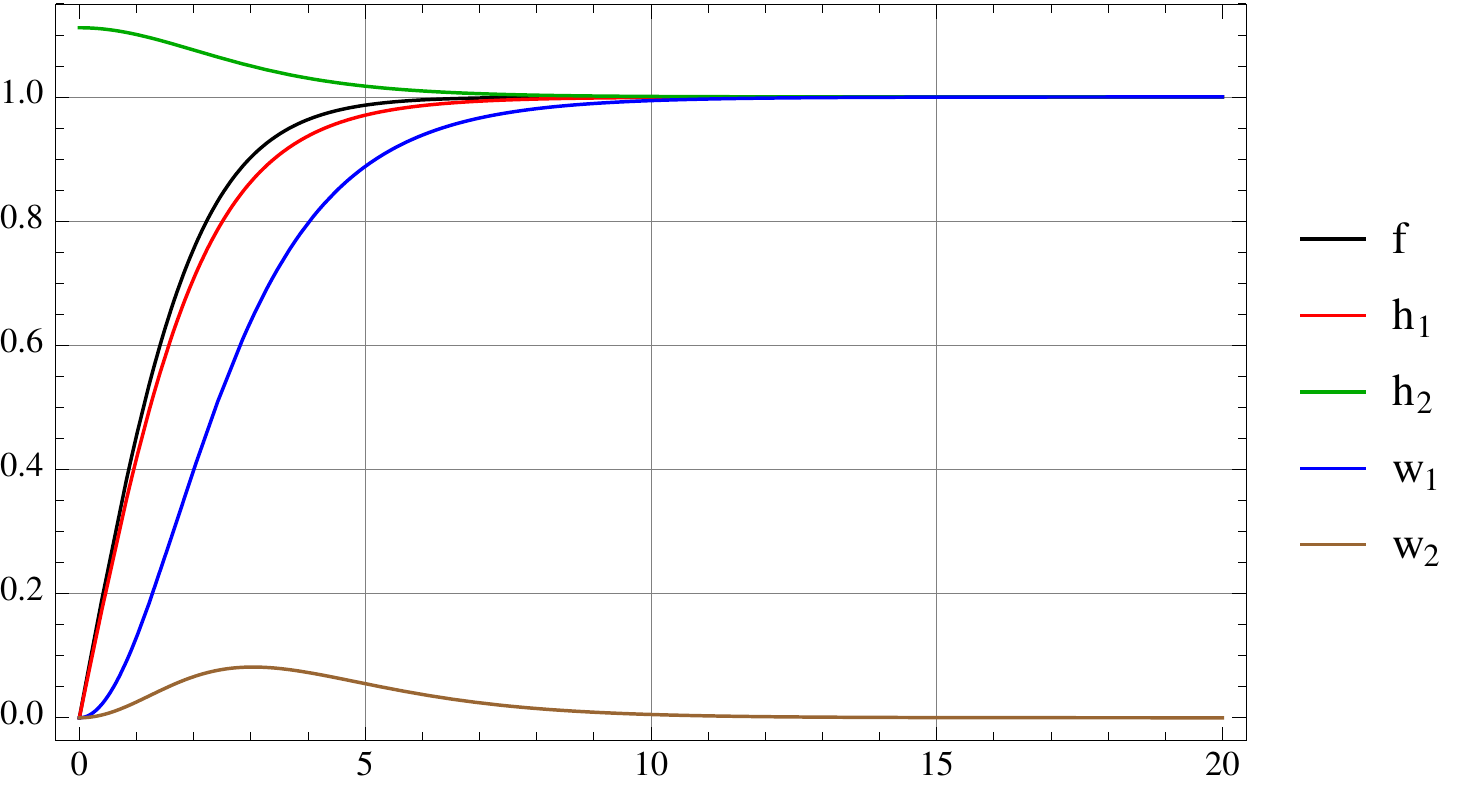}
\caption{Profile functions for the non-BPS DNALVS with 
$\tilde\omega=0.8$. Those for BPS NALVS with $\tilde \omega = 0$ are given in Fig.~\ref{fig:BPS_NALVS}.}
\label{fig:confs}
\end{center}
\end{figure}
Furthermore,  the conserved Noether charge density is given by
\beq
{\cal Q}^a = 2 i \Tr\left[
\left((\D^0 H)^\dagger H 
- H^\dagger \D^0 H \right)T^a
\right]
= \omega v^2\left( h_1^2 + h_2^2 - 2f h_1 h_2\right) \delta^{a3}.
\eeq
Thus, 
the tension and the Noether charge per unit length are given by
\beq
T(\tilde\omega) &=& 2\pi v^2 \int_0^\infty d\rho\ \rho 
\left[ {\cal K}_{\rm t} + {\cal K}_{\rm s} + {\cal V}\right],
\eeq
and
\beq
Q^a(\tilde \omega) = \int dx^1dx^2\ {\cal Q}^a = \frac{2\pi v^2}{\mu} \tilde \omega  \delta^{a3} \int d\rho\ \rho 
\left( h_1^2 + h_2^2 - 2f h_1 h_2\right).
\eeq

Before solving the equations of motion, 
let us first study stability of asymptotic state at $r \to \infty$. From Eq.~(\ref{eq:bc_infty}),
we perturb the fields as $h_{1,2} = 1 - \delta h_{1,2}$, $w_1 = 1 - \delta w_1$, $w_2 = \delta w_2$
and $f=1-\delta f$. Then the linearized equations for $\delta h_1$ and $\delta h_2$ read
\beq
\nabla^2
\left(
\begin{array}{c}
\delta h_1\\
\delta h_2
\end{array}
\right) = {\cal M}^2 
\left(
\begin{array}{c}
\delta h_1\\
\delta h_2
\end{array}
\right),\quad
{\cal M}^2 =
\left(
\begin{array}{cc}
\mu^2 - \frac{\omega^2}{2} & \frac{\omega^2}{2}\\
\frac{\omega^2}{2} & \mu^2 - \frac{\omega^2}{2}
\end{array}
\right).
\label{eq:fluctuation}
\eeq
The eigenvalues of the mass square matrix are $\mu^2$ and $\mu^2 - \omega^2$.
Thus, the fluctuations around configurations with $\omega^2 > \mu^2$ become tachyonic, so that
we have to set $|\omega| < \mu$, namely $|\tilde \omega|$ should be less than 1
as mentioned before.

We are now ready to solve the equations of motion. Since no analytic solutions can be obtained,
we solve them numerically. 
A numerical solution is shown in Fig.~\ref{fig:confs}.
As can be seen from Figs.~\ref{fig:BPS_NALVS} and \ref{fig:confs}, DNALVS with non-zero $\omega$ becomes fatter than 
the non-dyonic NALVS.
This is consistent with Eq.~(\ref{eq:fluctuation}) that 
leads to an asymptotic behavior $e^{-\sqrt{\mu^2 - \omega^2}\,r}$
in the profile functions.

In Sec.~\ref{sec:lorentz boost}, we have considered the spiral extension of the BPS DNALVS
by boosting the solution along the string axis. Obviously, similar spiral solution 
for the non-BPS DNALVS can be easily constructed.

\subsection{The non-BPS tension formula}

\begin{figure}[t]
\begin{center}
\includegraphics[width=16.5cm]{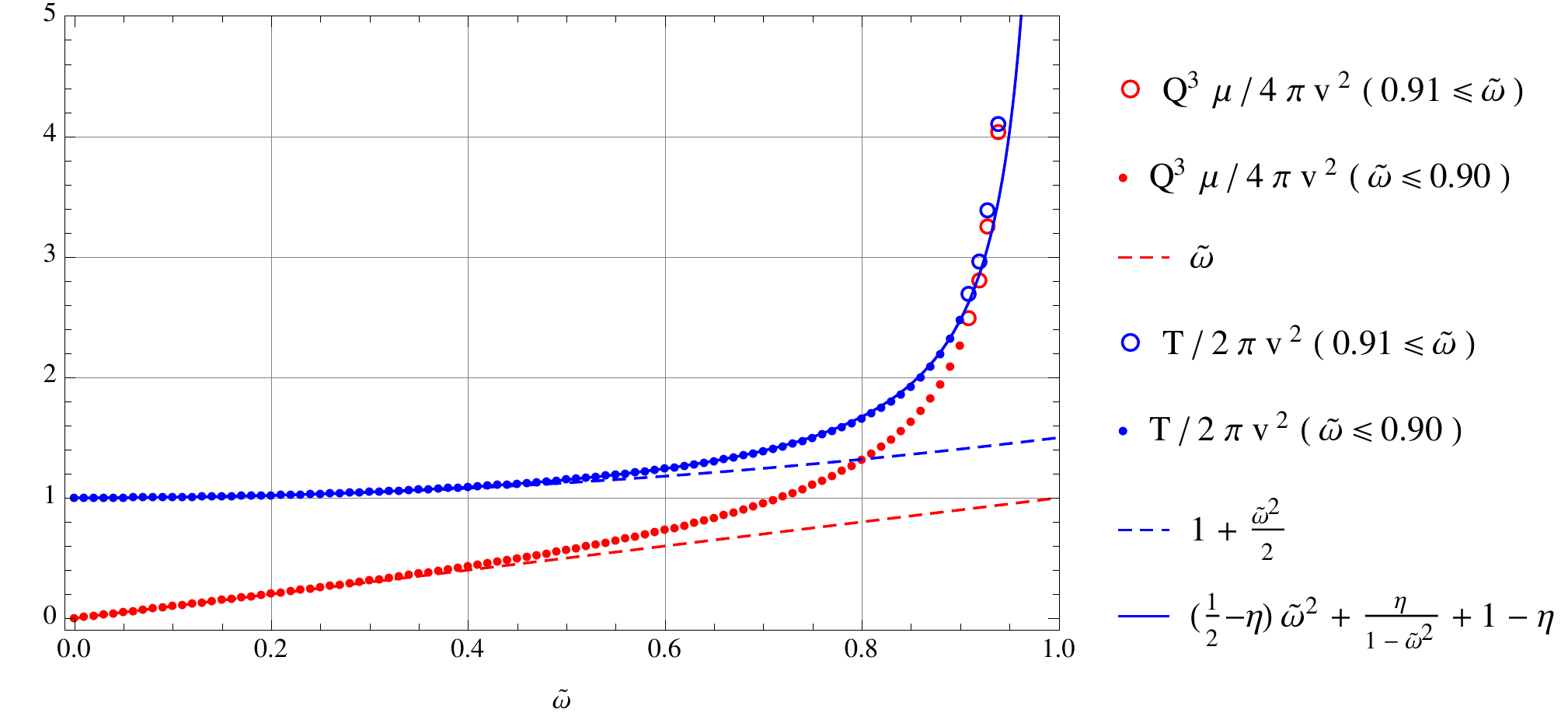}
\caption{The $\tilde \omega$ dependence of $T$ and $Q^3$ for $k=1$ DNALVS are shown. 
The dots are numerical results for $0 \le \tilde \omega \le 0.9$ and the red and blue circles are for
$0.91 \le \tilde \omega \le 0.94$. 
The blue dashed line corresponds to
$\frac{T}{2\pi v^2} = 1+\tilde\omega^2/2$ while the red dashed line corresponds to 
$\frac{\mu Q^3}{4\pi v^2} = \tilde \omega$. The blue solid curve stands for $T(\tilde\omega)$ given
in Eq.~(\ref{eq:global}).}
\label{fig:EQW}
\end{center}
\end{figure}

Let us next study the tension of DNALVS.
In Fig.~\ref{fig:EQW}, we show $\tilde \omega$-dependence in $T$ and $Q^3$ for 
$\tilde\omega = [0,0.94]$ with a step $\delta \tilde\omega = 0.01$.
For small $\tilde\omega$, 
we find that the numerical results are well fitted by the following functions, see Fig.~\ref{fig:EQW} 
\beq
\frac{T}{2\pi v^2} = 1 + \frac{1}{2}\tilde\omega^2 + {\cal O}(\tilde\omega^3),\qquad 
\frac{\mu Q^{3}}{4\pi v^2} =   \tilde \omega + {\cal O}(\tilde\omega^2).
\label{eq:num_MQ_U2}
\eeq
Note that, since we have the  identification $|q^3| = |Q^3|/2$, the second equation is consistent with Eq.~(\ref{eq:Phi_nonBPS}) 
obtained from the low energy effective action
with $m = 0$.
From these, we find a direct relation of $T$ and $Q^3$ for small $|Q^3|$, 
\beq
T = 2\pi v^2 + \frac{1}{4\beta} (Q^{3})^2 + \cdots.
\label{eq:T_quadratic}
\eeq
This is also consistent with the observation Eq.~(\ref{eq:tension_nBPS}).

\begin{figure}[t]
\begin{center}
\includegraphics[width=17cm]{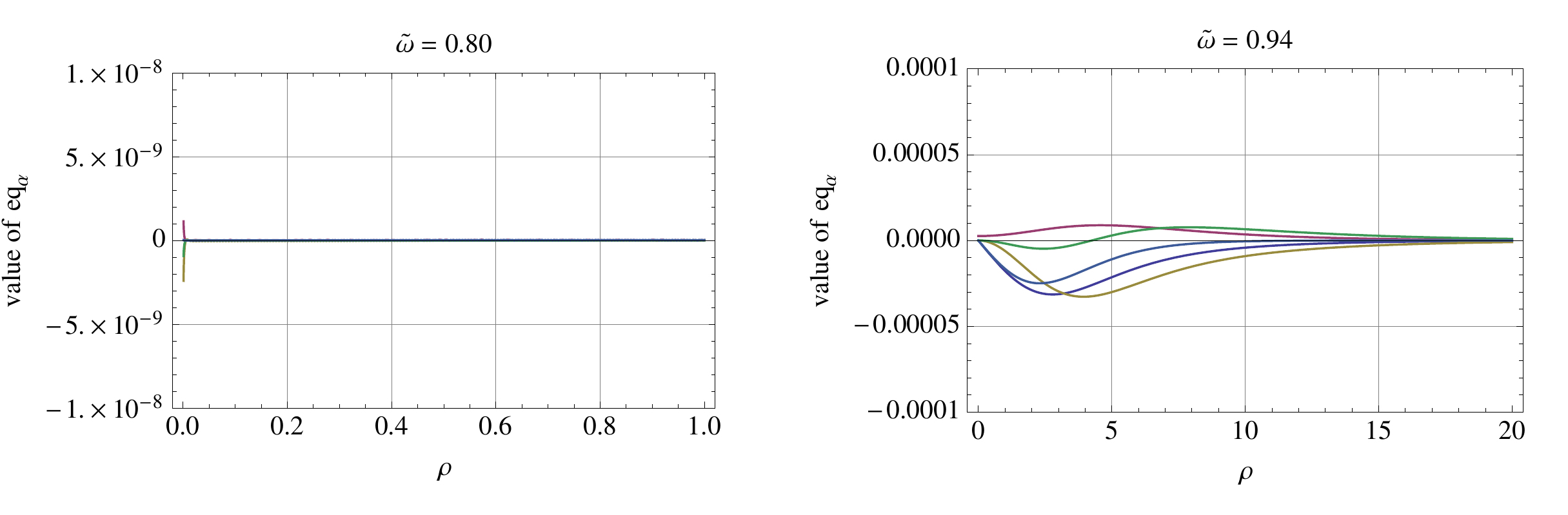} 
\caption{Numerical errors, namely values of ${\rm eq}_\alpha$ at the end of integration are shown.
The left panel is for $\tilde \omega = 0.80$ and the right one is for $\tilde \omega = 0.94$.}
\label{fig:errors}
\end{center}
\end{figure}

Remember that the effective theory description is valid only for sufficiently 
small $\tilde \omega$. On the other hand,
since we have solved the full equations of motion, we are now able
to go beyond the effective theory. Namely, it may be possible to find an appropriate function which 
can reproduce the numerical data not only for a small $\tilde \omega$ but also for a large $\tilde \omega$.
Such function should be a function of $\tilde \omega^2$ and singular at $\tilde \omega =1$.
Indeed, we find that the following function fits the numerical data quite well, see Fig.~\ref{fig:EQW},
\beq
\frac{T}{2\pi v^2} = 1 + 
\left(\frac{1}{2}-\eta\right) \tilde \omega^2 + \frac{\eta\,\tilde\omega^2}{1 - \tilde\omega^2}, \qquad
\eta = 0.311515.
\label{eq:global}
\eeq
This is somewhat surprising because the almost all numerical data are fitted by the function 
with only one parameter $\eta$. Two comments are in order.
Firstly, the expression (\ref{eq:global}) can be expanded in terms of $\tilde \omega$ as
\beq
\frac{T}{2\pi v^2} = 1 + \frac{\tilde\omega^2}{2} + \eta \left( \tilde \omega^4 + \tilde \omega^6 +  \cdots \right).
\label{eq:quartic}
\eeq
The first two terms are independent of $\eta$, and they are consistent with Eq.~(\ref{eq:num_MQ_U2}).
Secondly, as can be seen in Fig.~\ref{fig:EQW}, while Eq.~(\ref{eq:global}) reproduces 
the numerical data below $\tilde \omega = 0.90$ quite well, 
the data in $0.91 \le \tilde\omega \le 0.94$ are not fitted very well.
As to the solutions for $\tilde \omega > 0.94$, we cannot succeed in obtaining physically meaningful 
configurations.
A reason seems to be related to a numerical method and its accuracy.
We numerically solve Eqs.~(\ref{eq:eom1}) -- (\ref{eq:eom5}) by the so-called relaxation method. Namely,
we first regard
$\{h_{1,2},w_{1,2},f\}$ as functions of $\rho$ and an artificial time $\tau$ and
add dissipative terms to the right hand side Eqs.~(\ref{eq:eom1}) -- (\ref{eq:eom5}) as 
\beq
{\rm eq}_\alpha = \frac{\p X_\alpha(\rho,\tau)}{\p \tau},\quad 
X_\alpha = (h_1,h_2,w_1,w_2,f).
\label{eq:relaxation}
\eeq
We integrate these in $\tau$ with a suitable initial configuration $X_\alpha(\rho,\tau=0)$ that
satisfies the boundary conditions given in Eqs.~(\ref{eq:bc0}) and (\ref{eq:bc_infty}).
After a while in $\tau$, the profile functions converge. Since we have $\p X_\alpha/\p \tau = 0$,
the converged profile functions $X_\alpha (\rho,\infty)$ solve the genuine equations of motion.
In this way, we get the numerical solutions of $k=1$ for $0 \le |\tilde \omega| \le 0.94$.
Unfortunately, we failed to get numerical solutions above $|\tilde\omega| > 0.94$. In this region,
what we get is not a configuration localized around the vortex center but expanding toward spatial infinity.
Indeed, when we increase the computational domain,
the numerical solution becomes just fatter. Namely, they are sensitive to the box size, and
we cannot accept them as genuine vortex solutions. 
Relatedly, we also observe that a numerical error for a larger $|\tilde\omega|$
tend to be larger than that for a smaller $|\tilde \omega|$, see Fig.~\ref{fig:errors} where the values of 
${\rm eq}_\alpha$ in Eqs.~(\ref{eq:eom1}) -- (\ref{eq:eom5}) at the end of numerical integration
in $\tau$ are shown. For the numerical solutions 
with $|\tilde \omega| \ge 0.91$, the numerical errors are not so small.
Therefore, we guess that discrepancy between the formula (\ref{eq:global}) and 
the numerical results for $|\tilde \omega| \ge 0.91$ is caused by the numerical errors.

\begin{figure}[t]
\begin{center}
\includegraphics[width=15cm]{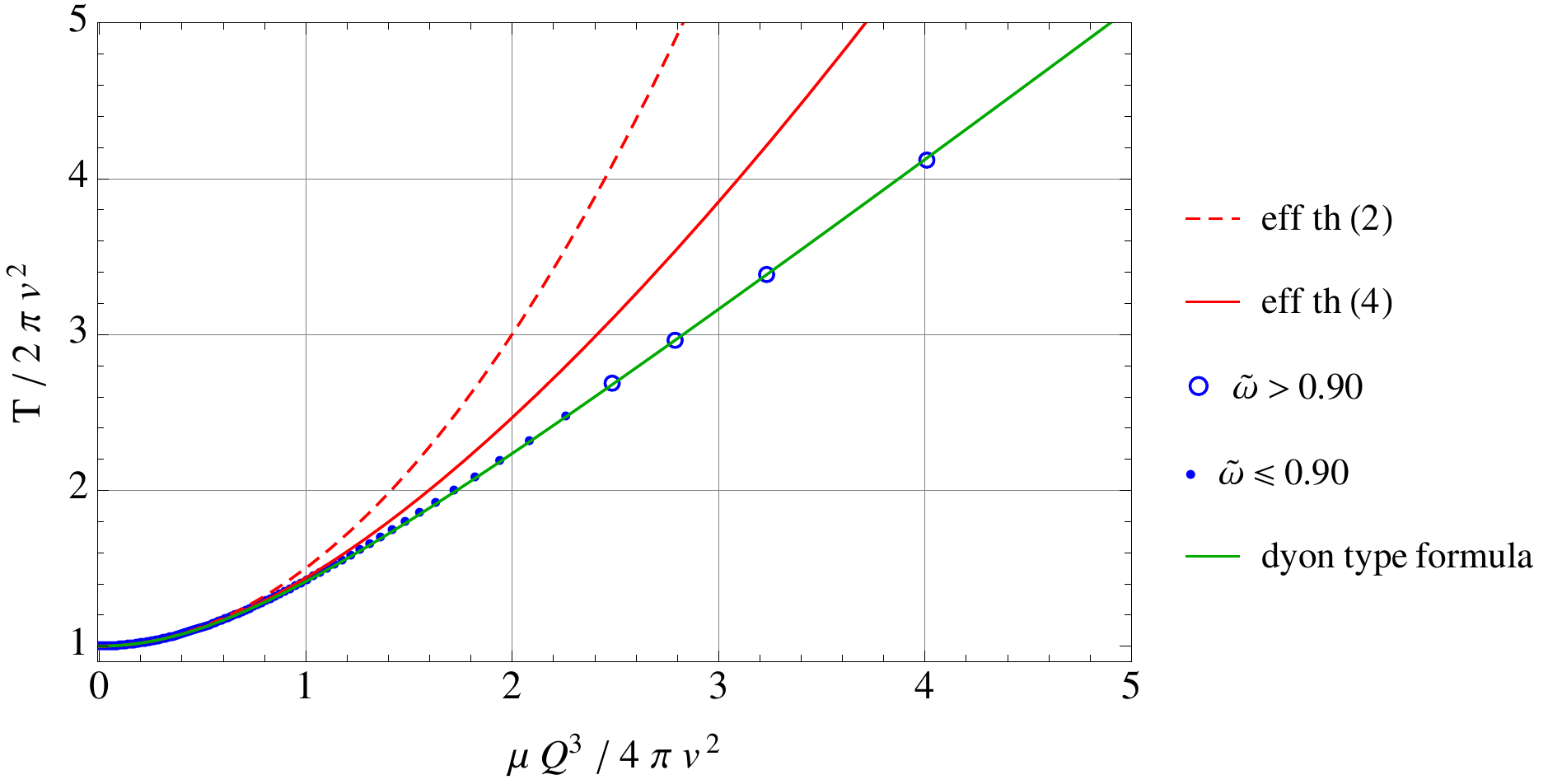} 
\caption{Comparison with the dyon-type formula (green line) given in Eq.~(\ref{eq:SUSY_mass_formula}) 
and results by
the effective field theories including terms up to quadratic (red-dashed line) and quartic (red line) 
derivative terms in
Eqs.~(\ref{eq:eff_q}) and (\ref{eq:eff_M}). The numerical data are the blue dots and the blue circles.}
\label{fig:eff_vs_num}
\end{center}
\end{figure}

Furthermore, we show a direct relation 
between the tension $T$ and the Noether charge $Q^3$ by regarding 
$\tilde \omega$ as a parameter,
see Fig.~\ref{fig:eff_vs_num}.
Surprisingly, we find that the dyon-type tension formula can reproduce the numerical result quite well, 
\beq
T_{k=1} \cong \sqrt{(2\pi v^2)^2 + \left(\frac{\mu}{2} Q^3\right)^2}.
\label{eq:SUSY_mass_formula}
\eeq
Here we have used the symbol $\cong$, in order not to forget the fact that the equality is not analytically
proven but is 
just verified by the numerical calculation. Anyway, the dyon-type formula works pretty well.
Actually, if we expand Eq.~(\ref{eq:SUSY_mass_formula}) to the quadratic order in $(\mu Q^3/2)/(2\pi v^2)$, we reproduce the result in Eq.~(\ref{eq:T_quadratic}).
%
We now encounter an unexpected result:
Remember, as is explained in Sec.~\ref{sec:2}, that the  tension formula (\ref{eq:BPS_dv_mass}) of 
the {\it BPS} DNALVS is of the dyonic-instanton type. This matches the conjecture 
that any solitons whose mass formulae are of the dyonic-instanton (dyon) 
type are BPS and their spatial co-dimensions are always even (odd).
Nevertheless, we have found that the tension of the {\it non-BPS} DNALVS
is approximated pretty well by
the dyon-type tension formula (\ref{eq:SUSY_mass_formula}).
To the best of our knowledge, no such non-BPS solitons have been known in the literature.
If the relation (\ref{eq:SUSY_mass_formula}) was really exact, this would have been 
the first counter-example to the conjecture.
However, as mentioned above, Eq.~(\ref{eq:SUSY_mass_formula}) is verified only by numerical calculations,
so we cannot conclude that the conjecture fails at this stage. Indeed,
in Sec.~\ref{sec:predict}, we will show that Eq.~(\ref{eq:SUSY_mass_formula}) cannot be exact. 
Therefore, the conjecture survives. We will propose another tension formula in Sec.~\ref{sec:predict}.

In addition to the minimally winding solution, we also 
examine axially symmetric solutions with higher winding numbers $k=2$ and 3. 
Since the dyonic solutions are non-BPS and have the Noether charge of the same sign, 
axially symmetric solution is very likely to be unstable. 
Despite of these, we assume an axial symmetry and solve the equations of motion (\ref{eq:eom1}) --
(\ref{eq:eom5}) with modifying them for $k=2,3$. 
Unexpectedly again, we find that the dyon-type formula approximately holds
\beq
T_{k\ge2} \approx \sqrt{(2\pi v^2k)^2 + \left(\frac{\mu}{2} Q^3\right)^2}.
\label{eq:dyon_nonBPS_higher}
\eeq
For small $Q^3$, this formula works well. However, increasing $Q^3$, the numerical data is
off the dyon-type formula. 
The coincidence between the numerical results and the formula (\ref{eq:dyon_nonBPS_higher})
is not as good as $k=1$ case especially for a large $\tilde \omega$, so we have used the symbol $\approx$ 
in Eq.~(\ref{eq:dyon_nonBPS_higher}).

\section{Higher derivative corrections to the low energy effective theory}
\label{sec:higher_derivatives}

\subsection{Quartic higher  derivative corrections}

Let us return to Eq.~(\ref{eq:quartic}).
The first term is the tension of the static non-Abelian vortex and the second term is the correction
of order ${\cal O}(\p_t^2)$ which can be explained from the view point of the low energy
effective theory in Sec.~\ref{sec:low energy}.  The third term is of order ${\cal O}(\p_t^4)$, and so it 
corresponds to a
higher derivative correction to  the effective Lagrangian given in Eq.~(\ref{eq:cp1}).

The higher derivative correction of the quartic order was obtained in Ref.~\cite{Eto:2012qda}
\beq
{\cal L}_{\rm eff}^{(2+4)} = \beta\left[
\frac{|\p_\alpha\phi|^2}{(1+|\phi|^2)^2} + \frac{\gamma}{\mu^2}\frac{|\p_\alpha\phi\p^\alpha \phi|^2}{(1+|\phi|^2)^4}
\right],
\label{eq:fourth_derivative}
\eeq
where $\gamma$ is a numerical constant defined by
\beq
\gamma \equiv \int d\rho\ \rho\left(1-h(\rho)^2\right)^2  \approx 0.830707.
\eeq
From this, the conserved charge and energy are given by 
\beq
q_{(2+4)}^3 &=& i\beta\left[
\frac{\phi \p_0\phi^*- \p_0\phi\phi^*}{(1+|\phi|^2)^2} + 
\frac{2\gamma}{\mu^2}\frac{\phi (\p_0\phi) (\p_0 \phi^*)^2 - \phi^*(\p_0 \phi^*)(\p_0\phi)^2}{(1+|\phi|^2)^4}
\right],
\label{eq:eff_q}\\
\delta T^{(2+4)} &=&  \beta\left[
\frac{|\p_0\phi|^2}{(1+|\phi|^2)^2} + \frac{3\gamma}{\mu^2}
\frac{|\p_0\phi|^4}{(1+|\phi|^2)^4}
\right].
\label{eq:eff_M}
\eeq
Note that, under the presence of the four derivative term, $\phi = e^{i\omega t}$ remains a solution
of the equations of motion. 
Plugging $\phi = e^{i\omega t}$ into these, the increment in the tension and the 
Noether charge per unit length are obtained as
\beq
\frac{\delta T^{(2+4)}}{2\pi v^2} &=& \frac{2}{\mu^2}\left[
\frac{\omega^2}{4} + \frac{3\gamma}{\mu^2}\frac{\omega^4}{16}
\right]
= 
\frac{1}{2}\tilde \omega^2 + \frac{3\gamma }{8}\tilde \omega^4,
\label{eq:tension_quartic}\\
\frac{\mu q_{(2+4)}^3}{2\pi v^2} &=& \frac{2}{\mu}\left[
\frac{\omega}{2} + \frac{\gamma}{\mu^2}\frac{\omega^3}{4}
\right] = \tilde\omega +  \frac{\gamma}{2}\tilde \omega^3.
\label{eq:q_quartic}
\eeq
Thus the coefficient of the quartic order reads
\beq
\frac{3\gamma }{8} = 0.311515.
\eeq
Comparing this with the coefficient of the third term of Eq.~(\ref{eq:quartic}), we should identify
\beq
\eta = \frac{3\gamma}{8},
\eeq
as is indeed taken in Eq.~(\ref{eq:global}).
In other words, we succeeded in computing the coefficient of the quartic derivative term only from
the tension formula, avoiding any complicated computations~\cite{Eto:2012qda}.
In Fig.~\ref{fig:eff_vs_num}, we show the numerical solutions, the dyonic-type 
tension formula (\ref{eq:SUSY_mass_formula}), and the results from effective action including
the quadratic and quartic derivative corrections.



\subsection{A prediction: All order higher derivative corrections}
\label{sec:predict}

Encouraged by success in finding the coefficient of the four derivative term from the tension
formula (\ref{eq:global}), we entertain hope to figure out higher derivative corrections of all order
to the effective theory for a single non-Abelian vortex.
To the best of our knowledge, higher derivative corrections only up to the quartic order have been
obtained in the literature. 
In order to simplify the problem, we consider DNALV in $2+1$ dimensions in this subsection. 
An advantage of this is that the effective theory is $0+1$ dimensional theory, so that there are less varieties 
for the higher derivative terms. For example, the $2n$-th order is simply proportional to $|\p_0 \phi|^{2n}$. 
On the contrary, there are several choices
for higher derivative terms in higher dimensions. For instance, there are two possibilities for four derivative
terms, 
$(\p_\alpha \phi\p^\alpha \bar\phi)^2$ and $|\p_\alpha \phi\p^\alpha \phi|^2$,
although only the latter appears in Eq.~(\ref{eq:fourth_derivative}).

Firstly, we expand Eq.~(\ref{eq:global}) and rewrite it in the following form 
\beq
T_{k=1} = 2\pi v^2 + \beta\left(\frac{\omega^2}{4} + \frac{\eta }{2}\sum_{n=2}^\infty \frac{\omega^{2n}}{\mu^{2n-2}}
\right).
\label{eq:expanded_tension}
\eeq
The second term should be compared with the effective theory with higher derivative corrections.
To this end, let us assume the effective Lagrangian to be the following form
\beq
{\cal L}_{\rm eff}^{(\infty)} = \beta \left( \frac{|\p_0\phi|^2}{(1+|\phi|^2)^2} 
+ \sum_{n=2}^\infty \frac{a_{2n}}{\mu^{2n-2}}
\frac{|\p_0\phi|^{2n}}{(1+|\phi|^2)^{2n}}\right),
\label{eq:eff_Lag_allorder}
\eeq
with $a_{2n}$ being real constants. Reflecting the symmetry of the original Lagrangian, 
this effective Lagrangian respects the $SU(2)_{\rm F}$ symmetry ($|\p_0\phi|^2/(1+|\phi|^2)^2$ is
an $SU(2)_{\rm F}$ invariant).
Note that, regardless of the coefficient $a_{2n}$, 
$\phi = e^{i\omega t}$ solves the equation of motion for this Lagrangian.
The corresponding Hamiltonian reads
\beq
H_{\rm eff}^{(\infty)} =  \beta \left( \frac{|\p_0\phi|^2}{(1+|\phi|^2)^2} 
+ \sum_{n=2}^\infty \frac{(2n-1)a_{2n}}{\mu^{2n-2}}
\frac{|\p_0\phi|^{2n}}{(1+|\phi|^2)^{2n}}\right).
\eeq
Plugging $\phi = e^{i\omega t}$ into the Hamiltonian,
we get
\beq
\delta T^{(\infty)} = \beta \left(\frac{\omega^2}{4} + \sum_{n=2}^\infty\frac{(2n-1)a_{2n}}{2^{2n}}\frac{\omega^{2n}}{\mu^{2n-2}}\right)
\eeq
Equating this with $T_{k=1} - 2\pi v^2$ in Eq.~(\ref{eq:expanded_tension}), we find the expansion 
coefficients $a_{2n}$ as
\beq
a_{2n} = \frac{2^{2n-1}}{2n-1} \, \eta.
\eeq
Plugging this back into Eq.~(\ref{eq:eff_Lag_allorder}), we end up a prediction to
the effective Lagrangian including higher derivative corrections of all order as
\beq
{\cal L}_{\rm eff}^{(\infty)} = 2\pi v^2 \left(\frac{1-2\eta}{2} X + \eta \sqrt{X} \tanh^{-1} \sqrt{X}\right),
\label{eq:all_order}
\eeq
where we have used $\sum_{n=1}^\infty X^n/(2n-1) = \sqrt{X}\tanh^{-1}\sqrt{X}$ and
\beq
X \equiv \frac{4|\p_0\phi|^2}{\mu^2(1+|\phi|^2)^2}.
\eeq

Our prediction fully depends on the global tension formula  (\ref{eq:global}) which we have verified 
only by numerical data. Since numerical errors are unavoidable in any numerical results, 
Eq.~(\ref{eq:global}) could not be exact but just an approximation.
If it is the case, the expression of the effective action given in Eq.~(\ref{eq:all_order}) would be modified
(Eq.~(\ref{eq:global}) is correct up to the quartic order).
However, this is not a big matter for us. What we would like to stress is that,
once somehow one gets the tension formula of DNALVS,
it is always possible to
derive an effective action including all order higher derivative corrections.  
Usually, obtaining the tension formula is much easier than
a straightforward but very complicated 
calculation for the higher derivative corrections. So, the tension formula helps us very much.

Regardless of the fact that the effective Lagrangian (\ref{eq:all_order}) is exact or approximate,
the effective Lagrangian (\ref{eq:all_order}) has an important property which the true effective 
action has to have: the true effective action should be singular at $\tilde \omega = 1$ since
DNALVS with $\tilde \omega =1$ does not exist. 
The effective theory with the quartic derivatives given in Eq.~(\ref{eq:fourth_derivative}) does not
have this property because $\tilde\omega = 1$ is beyond the scope.
On the other hand, the effective Lagrangian (\ref{eq:all_order}) is actually singular at $X = 1$.
Note that this occurs at $\tilde \omega = 1$ because of
\beq
X\big|_{\phi = e^{i\omega t}} = \tilde \omega^2.
\eeq

Remember the fact that, when we derive the effective action (\ref{eq:all_order}), we 
used only the tension formula (\ref{eq:global}) but we have not used any informations about 
the Noether charge. Note that, since
the Lagrangian (\ref{eq:all_order}) is invariant under the $U(1)$ transformation $\phi \to e^{i\alpha}\phi$,
the Lagrangian (\ref{eq:all_order}) has its own Noether charge. For the solution $\phi = e^{i\omega t}$, 
the tension and the Noether charge per unit length for the Lagrangian (\ref{eq:all_order}) are given by
\beq
\frac{\delta T_{(\infty)}}{2\pi v^2} &=& 
\left(\frac{1}{2}-\eta\right) \tilde \omega^2 + \frac{\eta\,\tilde\omega^2}{1 - \tilde\omega^2},
\label{eq:all_tension}\\
\frac{\mu q^3_{(\infty)}}{2\pi v^2} &=& 2 \left(\frac{1}{2} - \eta \right) \tilde\omega + \frac{\eta\,\tilde\omega}{1-\tilde\omega^2}
+\eta  \tanh ^{-1}\tilde\omega.
\label{eq:all_q}
\eeq
By definition, Eq.~(\ref{eq:all_tension}) coincides with Eq.~(\ref{eq:global}). Of course, if we expand 
Eq.~(\ref{eq:all_tension}) in $\tilde\omega$, it also reproduces the result (\ref{eq:tension_quartic}) from the
effective action including the quartic derivatives.
Similarly, if $q^3_{(\infty)}$ is expanded in $\tilde\omega$, we have
\beq
\frac{\mu q^3_{(\infty)}}{2\pi v^2} = \tilde\omega + \frac{4\eta}{3}\tilde\omega^3 + 
\frac{6\eta}{5}\tilde\omega^5 + \cdots,
\label{eq:expand_q_all}
\eeq
which is again consistent with (\ref{eq:q_quartic}).
From Eqs.~(\ref{eq:all_tension}) and (\ref{eq:all_q}), one can in principle derive a direct
relation beween $T_{(\infty)}/2\pi v^2= 1 + \delta T_{(\infty)}/2\pi v^2$ and $\mu q^3_{(\infty)}/2\pi v^2$.
But in practice it is not easy to express $T_{(\infty)}$ as a simple function of $\mu q^3_{(\infty)}$.
At least it is obviously different from the dyon-type tension formula (\ref{eq:SUSY_mass_formula}).
Now a question is which is more plausible, the dyon-tension formula (\ref{eq:SUSY_mass_formula})
or the relation from Eqs.~(\ref{eq:all_tension}) and (\ref{eq:all_q}).

 \begin{figure}[t]
\begin{center}
\includegraphics[width=15cm]{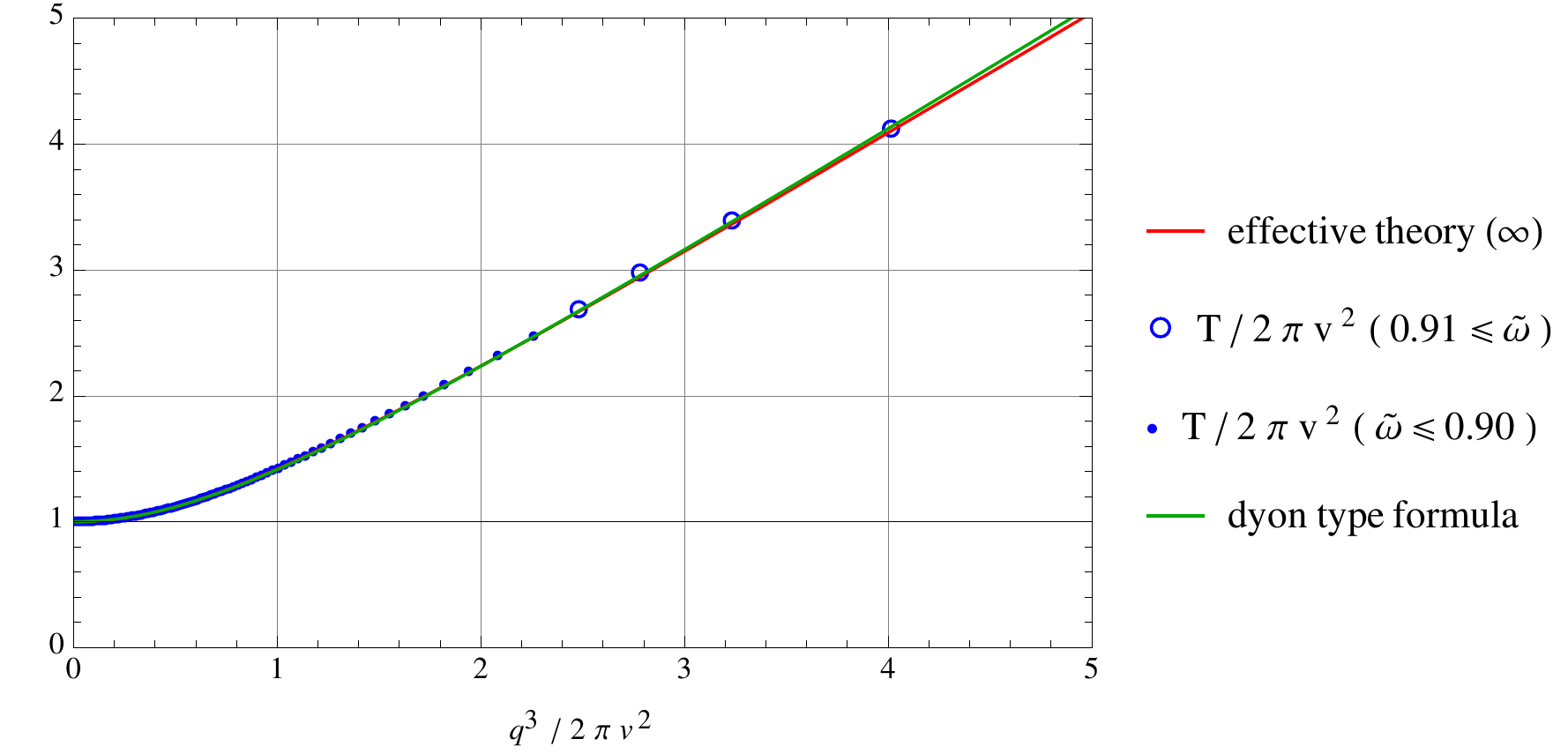} 
\caption{Comparison with the dyon-type formula (green line) given in Eq.~(\ref{eq:SUSY_mass_formula}) 
and the relation between $T$ and $Q^3$ derived from Eqs.~(\ref{eq:all_tension}) and (\ref{eq:all_q}).
The dots are the numerical results.}
\label{fig:QvsT_all_order}
\end{center}
\end{figure}

In order to verify validity of the former, let us assume that the global tension formula (\ref{eq:global}) and
the dyon-type tension formula (\ref{eq:SUSY_mass_formula}) are correct. Then, the $\tilde\omega$ dependence
in $q^3$ reads
\beq
\frac{\mu q^3}{2\pi v^2} &=& \sqrt{\left(
1 + 
\left(\frac{1}{2}-\eta\right) \tilde \omega^2 + \frac{\eta\,\tilde\omega^2}{1 - \tilde\omega^2}
\right)^2-1} \non
&=& \tilde\omega + \left(\frac{1}{8} + \eta \right)\tilde\omega^3 
+ \frac{1}{128} \left(-64 \eta ^2+176 \eta -1\right)\tilde\omega^5
+ \cdots.
\label{eq:expand_q_dyon}
\eeq
The difference from Eq.~(\ref{eq:expand_q_all}) 
appears at the order $\tilde \omega^3$. Since this is inconsistent with 
the expression (\ref{eq:q_quartic}) obtained from the effective Lagrangian, the 
dyon-type tension formula (\ref{eq:SUSY_mass_formula}) cannot be exact but just an appropriate expression.
Nevertheless, it is still surprising that the dyon-type tension formula  (\ref{eq:SUSY_mass_formula})
approximates the numerical data pretty well, see Fig.~\ref{fig:eff_vs_num}.
This coincidence can be also seen in the coefficients of Eqs.~(\ref{eq:expand_q_all}) and (\ref{eq:expand_q_dyon});
$4\eta/3 = 0.415353$, $1/8 + \eta = 0.436515$, 
$6\eta/5 = 0.373818$,
$\left(-64 \eta ^2+176 \eta -1\right)/128 = 0.372000$, and so on.
Furthermore, one can easily check the following 
\beq
\frac{d(T_{(\infty)}/2\pi v^2)}{d(\mu q^3_{(\infty)}/2\pi v^2)} 
= \frac{dT_{(\infty)}}{d\tilde\omega}\frac{d\tilde\omega}{d(\mu q^3_{(\infty)})}
= \tilde\omega
\to \left\{
\begin{array}{ccl}
\dfrac{\mu q^3}{2\pi v^2} & & \text{for}\quad \tilde\omega \to 0\\
&&\\
1 & & \text{for}\quad \tilde\omega \to 1
\end{array}
\right..
\eeq
This asymptotic behavior is consistent with that for the dyon-type tension formula (\ref{eq:SUSY_mass_formula}).
Therefore, Eqs.~(\ref{eq:all_tension}) and (\ref{eq:all_q}) gives us an inplicit function $T(Q^3)$
which is almost degenerate with the dyon-type tension formula $T(Q^3) = \sqrt{(2\pi v^2)^2 + (\mu Q^3/2)^2}$.
In Fig.~\ref{fig:QvsT_all_order}, we show the numerical data, the dyon-type tension formula (\ref{eq:SUSY_mass_formula}), and the relation from Eqs.~(\ref{eq:all_tension}) and (\ref{eq:all_q}).
Indeed, it is hard to find differences among these three below $\tilde \omega \le 0.90$.

In summary, we conclude that
the dyon-type tension formula (\ref{eq:SUSY_mass_formula}) is approximate relation, and
the more plausible tension formula is implicitly given by Eqs.~(\ref{eq:all_tension}) and (\ref{eq:all_q})
which are derived from the low energy effective Lagrangian (\ref{eq:all_order}).

\section{Dyonic non-Abelian semi-global vortex strings}
\label{sec:massless_dyonic_SGV}

In this section, we are going to study the dyonic non-Abelin vortex strings not in the $U(\NC)$ Yang-Mills-Higgs
model but in the $SU(\NC)$ Yang-Mills-Higgs model. A difference is  just ungauging the overall $U(1)$ symmetry
of the model in the previous sections. However, this drastically changes properties of the non-Abelian vortex strings.
The distinction stands out even in the model with $\NC=1$. In the $U(1)$ gauge theory (the Abelian-Higgs theory),
vortices  are  the so-called local vortex strings (or the Nielsen-Olesen vortex strings) 
whose tensions are finite.
On the other hand, strings in a global model (no gauge symmetries) are the so-called global vortex strings.
In contrast, tensions of the global vortices are logarithmically divergent.
In addition, no global vortices as BPS states have been found in the literature. For $\NC>1$ case in which
only $SU(\NC)$ is gauge symmetry and the overall $U(1)$ is global symmetry, 
the same can be said. Their tensions are logarithmically divergent and they are always
non-BPS states. They are sometimes called the semi-superfluid vortex strings especially in the context
of the high density QCD in which
they were firstly found~\cite{Balachandran:2005ev}. 
In this way, the non-Abelian strings in $U(\NC)$ gauge theory and $SU(\NC)$ gauge theory are quite different.
Nevertheless, they share an important property that they have normalizable non-Abelian orientational moduli.
Hence, as in the $U(\NC)$ case, the non-Abelian vortex strings in $SU(\NC)$ gauge theories can be
extended to dyonic ones.
As explained in the Introduction, we prefer the non-Abelian semi-global
vortex string (NASGVS) rather than the semi-superfluid string.
Please do not confuse it with the other non-Abelian vortex strings:
the non-Abelian local vortex string~\cite{Hanany:2003hp,Auzzi:2003fs,Shifman:2004dr,Hanany:2004ea}, 
the non-Abelian semi-local vortex string~\cite{Hanany:2003hp,Shifman:2006kd,Eto:2007yv,Auzzi:2008wm}, and
the non-Abelian global vortex string~\cite{Nitta:2007dp,Nakano:2007dq,Eto:2009wu,Eto:2013bxa}.

In order to study a dyonic NASGVS (DNASGVS),
we slightly change the bosonic Lagrangian of $\N=2$ supersymmetric Yang-Mills-Higgs Lagrangian
given in Eq.~(\ref{eq:lag_U(N)}). Firstly, we ungauge the overall $U(1)$ symmetry. Then,
we discard unimportant fields $\Sigma$ and $H^{i=2}$ from the beginning,
and set $M=0$.
As result, we start with the following simple Lagrangian 
\beq
\Lag_{SU(\NC)} = \Tr\bigg[
-\frac{1}{2g^2}\hat F_{\mu\nu}\hat F^{\mu\nu} 
+ \hat \D_\mu H (\hat \D^\mu H)^\dagger 
- \frac{g^2}{4}\left(HH^\dagger - v^2{\bf 1}_{\NC}\right)^2
\bigg],
\label{eq:Lag_SU}
\eeq
where $\hat F_{\mu\nu}$ and $\hat \D_\mu$ are the field strength and covariant derivative for
the $SU(\NC)$ gauge field.
Note that this Lagrangian cannot be embedded into any supersymmetric models.
This is because the overall $U(1)$ symmetry is not gauge symmetry, so that it is impossible to introduce 
the term proportional to unit matrix [$D_{U(1)} = (\Tr[HH^\dagger]/\NC - v^2){\bf 1}_{\NC}$] in the scalar potential. 
The symmetry of the model is 
\beq
G = SU(\NC)_{\rm C} \times U(1)_{\rm B} \times SU(\NC)_{\rm F}.
\eeq
The first $SU(\NC)_{\rm C}$ is the gauge symmetry while the rest symmetry
$U(1)_{\rm B} \times SU(\NC)_{\rm F}$ is the global symmetry.
The vacuum moduli space is $S^1$ which is parametrized by 
\beq
H = v e^{i\varphi} {\bf 1}_{\NC}.
\eeq
The phase $e^{i\varphi}$ corresponds to the spontaneously broken $U(1)_{\rm B}$.
At any points in the vacuum moduli space, the symmetry $G$ is spontaneously broken as
\beq
SU(\NC)_{\rm C} \times U(1)_{\rm B} \times SU(\NC)_{\rm F}
\quad \to \quad
SU(\NC)_{\rm C+F}.
\eeq
The spectra in the Higgs vacuum split into two masses; the one is
$\mu^2 = g^2v^2$ for the $SU(\NC)$ gauge field ($\NC^2-1$ degrees of freedom) and for
the real part of $H$ ($\NC^2$ degrees of freedom), 
and the other is a Nambu-Goldstone zero mode associated with
the spontaneously broken $U(1)_{\rm B}$ symmetry (the phase of the trace part of $H$).
The massless mode exists 
alone because of the absence of the $U(1)$ gauge field.

\begin{figure}[t]
\begin{center}
\includegraphics[width=12cm]{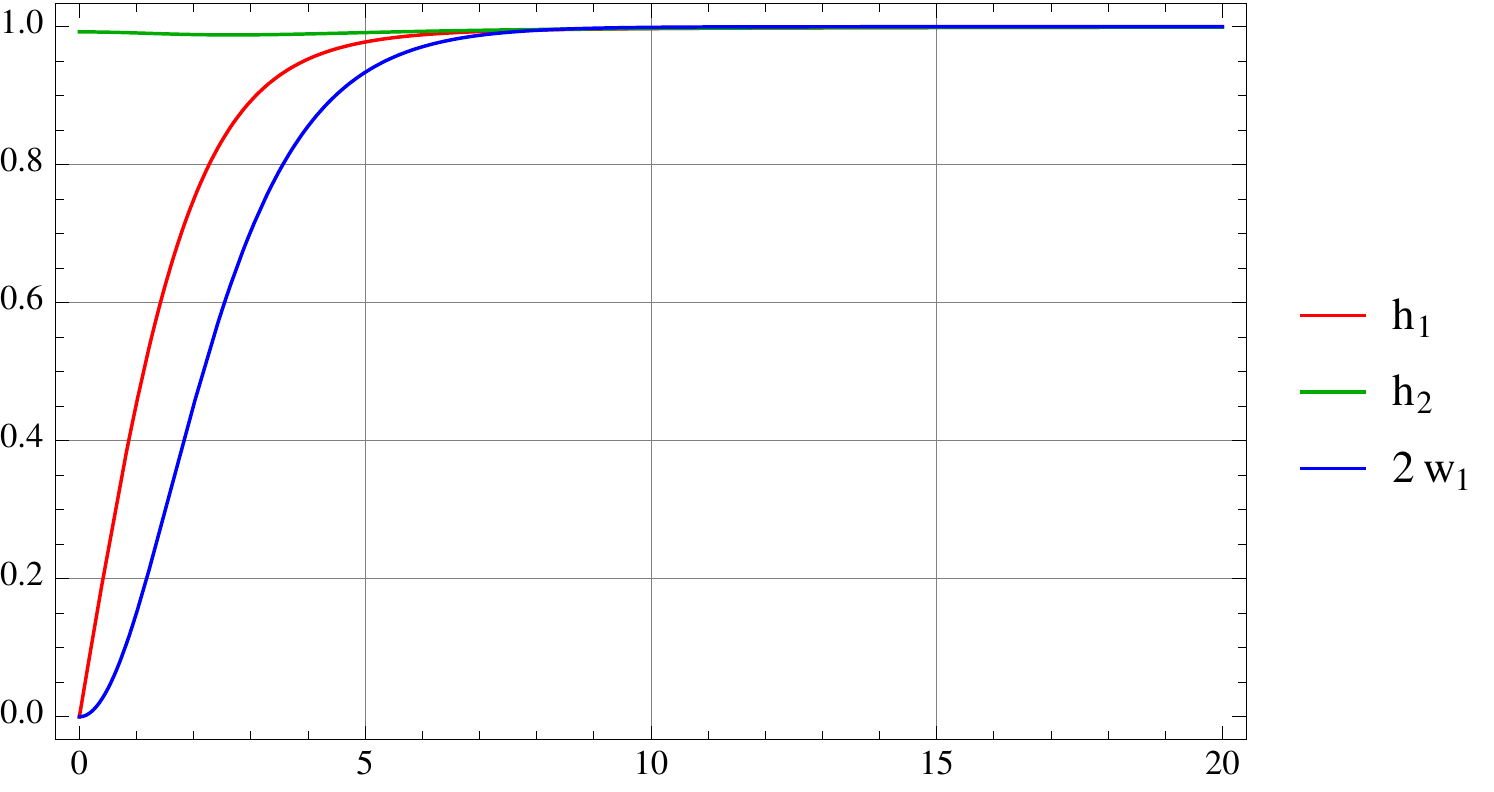}\\\ \\
\includegraphics[width=12cm]{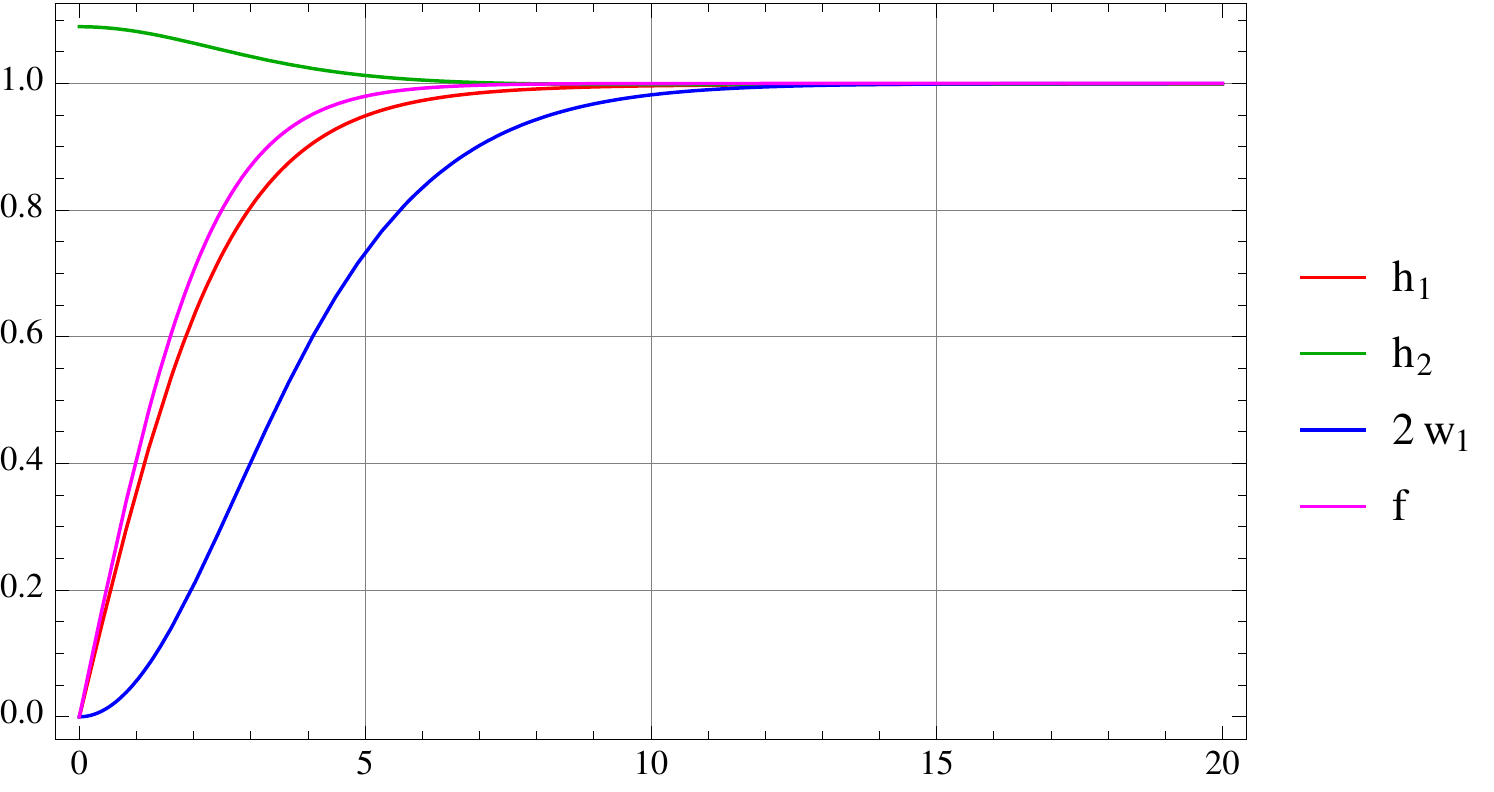}
\caption{Profile functions for a static semi-global non-Abelian vortex with 
$\tilde\omega=0$ (the upper panel) and for a dyonic semi-global non-Abelian vortex with
$\tilde\omega=0.7$ (the bottom panel).}
\label{fig:conf_SGV}
\end{center}
\end{figure}
Since neither NASGVS nor DNASGVS is a BPS state, we have to solve
the equations of motion,
\beq
\hat\D_\mu\hat\D^\mu H + \frac{g^2}{2}(HH^\dagger - v^2{\bf 1}_2)H = 0,
\label{eq:SGV_eom_1}\\
\frac{1}{g^2}\hat \D_\mu\hat F^{\mu\nu} + i\left<H\left(\hat \D^\nu H\right)^\dagger
- \left(\hat\D^\nu H\right) H^\dagger \right> = 0,
\label{eq:SGV_eom_2}
\eeq
where $\left<X\right>$ stands for traceless part of an $\NC\times\NC$ matrix $X$.
In what follows, we will consider the case of $\NC = 2$ for simplicity.
For a static (non-dyonic) configuration, we set $\p_{0,3} = 0$ and $W_{0,3}=0$, and 
make the following Ansatz~\cite{Eto:2009kg}
\beq
H &=&  v\left(
\begin{array}{cc}
h_1(r)e^{i\theta} & 0 \\
0 & h_2(r)
\end{array}
\right),
\label{eq:ansatz_H}\\
W_1 + i W_2 &=&
- \frac{ie^{i\theta}}{r} \left(
\begin{array}{cc}
w_1(r) & 0\\
0 & -w_1(r)
\end{array}
\right)
.
\label{eq:ansatz_W}
\eeq
Plugging these into Eqs.~(\ref{eq:SGV_eom_1}) and (\ref{eq:SGV_eom_2}), we get
the three 2nd order differential equations for $h_{1,2}$ and $w_1$.
In terms of the dimensionless coordinate $\rho = \mu r$, they are expressed as
\beq
h_1''+\frac{h_1'}{\rho}-\frac{(1-w_1)^2}{\rho^2}h_1
-\frac{1}{2} \left(h_1^2-1\right)h_1 =0,
\label{eq:eomSU_1}\\
h_2''+\frac{h_2'}{\rho}-\frac{w_1^2}{\rho^2}h_2-\frac{1}{2}(h_2^2-1)h_2 = 0,
\label{eq:eomSU_2}\\
w_1'' - \frac{w_1'}{\rho}+\frac{1}{2}\left(h_1^2(1-w_1)-h_2^2w_1\right)= 0.
\label{eq:eomSU_3}
\eeq
We solve these with the boundary conditions
\beq
h_1 \to 1,\quad
h_2 \to 1,\quad
w_1 \to \frac{1}{2},\qquad &{\rm as}&\quad \rho\to\infty,\\
h_1 \to 0,\quad
h_2' \to 0,\quad
w_1 \to 0,\qquad &{\rm as}&\quad \rho\to0.
\eeq
It is hard to analytically solve the above differential equations but is easy to 
numerically solve them. A numerical solution is shown in the upper panel of Fig.~\ref{fig:conf_SGV}.
Note that $h_2$ looks like uniformly 1 in Fig.~\ref{fig:conf_SGV}, but it is slightly different from 1
near the origin. Indeed,
$h_2=1$ is not a solution of Eq.~(\ref{eq:eomSU_2}), if $w_1 \neq 0$.

Let us study asymptotic behaviors of the non-dyonic NASGVS by perturbing the fields 
far from the origin as
\beq
h_1 = 1 - \delta h_1,\quad h_2 = 1 - \delta h_2,\quad w_1 = \frac{1}{2} - \delta w_1.
\label{eq:perturbe}
\eeq
The linearized equations at $r\to\infty$ for trace and traceless parts
\beq
\delta F \equiv \delta h_1 + \delta h_2,\qquad 
\delta G \equiv \delta h_1 - \delta h_2
\label{eq:linear_FG}
\eeq
are given by
\beq
\left(\triangle  - 1  \right) \delta F = -\frac{1}{2\rho^2},\qquad
\left(\triangle  - 1  \right) \delta G = 0,
\eeq
and that for $\delta w_1$ is given by
\beq
\left(\triangle'  - 1  \right) \delta w_1 = -\frac{1}{2}\delta G,
\label{eq:dw1}
\eeq
where we defined $\triangle = \frac{d^2}{d\rho^2} + \frac{1}{\rho}\frac{d}{d\rho}$ and 
$\triangle' = \frac{d^2}{d\rho^2} - \frac{1}{\rho}\frac{d}{d\rho}$.
Eq.~(\ref{eq:dw1}) can be cast into a standard form
\beq
\left(\triangle  - 1  \right) \frac{\delta w_1}{\rho} = 0.
\eeq
Solving these equations, we find
\beq
\delta F &=& \frac{1}{2\rho^2} + {\cal O}\left(\rho^{-4}\right),
\label{ref:dF}\\
\delta G &=& q_{\rm s} K_0(\rho) + {\cal O}\left((e^{-\rho})^2\right),\\
\delta w_1 &=& q_{\rm g} \rho K_0(\rho) + {\cal O}\left((e^{-\rho})^2\right),
\eeq
where $K_n(\rho)$ is the modified Bessel functions of the second kind and $q_{\rm s}$ and $q_{\rm g}$ are
numerical constants. While the Higgs mechanism for the $SU(\NC)$ part ensures the exponentially
small tails for $\delta G$ and $\delta w_1$, the trace part $\delta F$ has a long tail of power behavior
due to the massless Nambu-Goldstone mode~\cite{Eto:2009kg}. 
Note that this long tail in the trace part makes tension logarithmically divergent.
This can be seen by looking at the kinetic term of $H$ 
\beq
\int d^2x\ \Tr[\hat \D_i H \hat \D_i H^\dagger]
\sim \int d^2x\ \frac{v^2}{2\rho^2} = \pi v^2 \int d\rho \frac{1}{\rho} = \pi v^2 \log L,
\eeq
where $L$ is a IR cut off in unit of $\mu^{-1}$.

Next we extend NASGVS to dyonic ones.
As in Sec.~\ref{sec:non_BPS_dyonic_U2_vortex}, we rotate the static Ansatz in 
Eqs.~(\ref{eq:ansatz_H}) and (\ref{eq:ansatz_W}) by the time-dependent $SU(2)_{\rm C+F}$ matrix
defined by Eq.~(\ref{eq:Ut})
\beq
H &=& \tilde U^\dagger(t) \left[ v\left(
\begin{array}{cc}
h_1(r)e^{i\theta} & 0 \\
0 & h_2(r)
\end{array}
\right)\right]
\tilde U(t),\\
\bar W &=& \tilde U^\dagger(t)
\left[
- \frac{ie^{i\theta}}{2r} \left(
\begin{array}{cc}
w_1(r) & 0\\
0 & -w_1(r)
\end{array}
\right)
\right]
\tilde U(t),\\
W_0 &=& \tilde U^\dagger(t)
\left[
\frac{\omega}{2}
\left(
\begin{array}{cc}
0 & 1-e^{i\theta}f(r) \\
1-e^{-i\theta}f(r) & 0
\end{array}
\right)
\right]\tilde U(t).
\eeq
Plugging these into the equations of motion (\ref{eq:SGV_eom_1}) and (\ref{eq:SGV_eom_2}), we find
\beq
h_1''+\frac{h_1'}{\rho}-\frac{(1-w_1)^2}{\rho^2}h_1
-\frac{1}{2} \left(h_1^2-1\right)h_1 = - \frac{\tilde \omega^2}{4} \left(\left(f^2+1\right) h_1-2 f h_2\right),\\
h_2''+\frac{h_2'}{\rho}-\frac{w_1^2}{\rho^2}h_2-\frac{1}{2}(h_2^2-1)h_2= - \frac{\tilde \omega^2}{4}
\left(-2fh_1+(1+f^2)h_2\right),\\
w_1'' - \frac{w_1'}{\rho}+\frac{1}{2}\left(h_1^2(1-w_1)-h_2^2w_1\right) = \frac{\tilde \omega^2}{2}(1-2w_1)f^2,\\
f''+\frac{f'}{\rho}-\frac{(1-2w_1)^2}{\rho^2}f-\frac{1}{2}\left((h_1^2+h_2^2)f-2h_1h_2\right) = 0,
\eeq
where we have again used $\rho = \mu r$ and $\tilde \omega = \omega/\mu$.
The boundary conditions for $h_1$, $h_2$ and $f$ are the same as those given in 
Eqs.~(\ref{eq:bc0}) and (\ref{eq:bc_infty}). The one for $w_1$ is given by
\beq
w_1(0) = 0,\quad w_1(\infty) = \frac{1}{2}.
\eeq
A numerical solution is shown in the lower panel in Fig.~\ref{fig:conf_SGV}.
As in the case of the $U(2)$ gauge theory, the dyonic configuration becomes fatter than the static configuration.
This fact also reflects in the asymptotic behavior. 
To see this, let us perturb the fields as Eq.~(\ref{eq:perturbe}) with $f = 1 - \delta f$. Then we have
the following linearized equations
\beq
\left(\triangle - 1\right) \delta F &=& - \frac{1}{2\rho^2},\\
\left(\triangle - (1-\tilde\omega^2)\right) \delta G &=& 0,\\
\left(\triangle' - (1-\tilde \omega^2)\right) \delta w_1 &=& - \frac{1}{2}\delta G,\\
\left(\triangle - 1\right) \delta f &=& 0.
\eeq
The equation for $\delta F$ is unchanged from Eq.~(\ref{eq:linear_FG}), so the asymptotic behavior
is the same as one given in Eq.~(\ref{ref:dF}). The equations for $\delta G$ and $\delta w_1$ are the same 
as those given in Eqs.~(\ref{eq:linear_FG}) and (\ref{eq:dw1}) except for the mass square being
replaced by $1 \to 1 -\tilde \omega^2$. Thus the asymptotic behaviors are
\beq
\delta G = q_{\rm s}' K_0\left(\sqrt{1-\tilde\omega^2}\rho\right),\quad
\delta w_1 = q_{\rm g}'\left(\sqrt{1-\tilde\omega^2}\rho\right) K_0\left(\sqrt{1-\tilde\omega^2}\rho\right),\quad
\delta f = q_0K_0(\rho),
\eeq
where $q_{\rm s}'$, $q_{\rm g}'$ and $q_0$ are numerical constants.
The quantity $\sqrt{\mu^2 - \omega^2}$ plays a role of an ``effective'' mass  for
the dyonic configurations, and so it should be positive definite. When it becomes negative, the configuration
becomes unstable since the effective mass is tachyonic. Thus, for stable configurations, we should
restrict
\beq
|\omega| < \mu,
\eeq
as in the $U(2)$ case in Sec.~\ref{sec:non_BPS_dyonic_U2_vortex}.

The Noether charge density is formally the same as that in the $U(2)$ case. 
\beq
{\cal Q}^a \equiv 2 i \Tr\left[
\left((\hat\D^0 H)^\dagger H 
- H^\dagger \hat\D^0 H \right)T^a
\right]
= \omega v^2 \left( h_1^2 + h_2^2 - 2f h_1 h_2\right) \delta^{a3}.
\eeq
Since the tension of NASGVS is logarithmically divergent, one may anticipate that
the Noether charge also diverges. But this is not the case. Indeed, it is exponentially small
at spatial inifnity
\beq
{\cal Q}^a \to 2\omega v^2 \delta f \delta^{a3} = 2\omega v^2 q_0 K_0(\rho) \delta^{a3},\quad
(\rho \to \infty),
\eeq
so that the Noether charge per unit length is finite,
\beq
Q^a = \int dx^1dx^2\ {\cal Q}^a.
\eeq
\begin{table}[t]
\begin{center}
\caption{The masses (in the unit of $2\pi v^2$) of 
a semi-global vortex as function of the frequency $\tilde \omega$ 
$(= 0,0.1,0.2,0.3,0.4,0.5,0.6)$ and the IR
cutoff scale $L$ $(=50,100,150)$ in the unit of $\mu^{-1}$.}
\ \\
\begin{tabular}{c|c:cccccc}
\hline
$L$ & $T(0;L)$ & $\delta T(0.1)$ & $\delta T(0.2)$ & $\delta T(0.3)$ 
& $\delta T(0.4)$ & $\delta T(0.5)$ & $\delta T(0.6)$\\
\hline
50   & 2.5456 & 0.0059 & 0.0243 & 0.0576 & 0.1110 & 0.1955 & 0.3356\\
100 & 2.8922 & 0.0059 & 0.0243 & 0.0576 & 0.1110 & 0.1955 & 0.3356\\
150 & 3.0949 & 0.0059 & 0.0243 & 0.0576 & 0.1110 & 0.1955 & 0.3356\\
\hline
\end{tabular}
\label{tab:1}
\end{center}
\end{table}
Hence, nevertheless the tension itself diverges, we expect that excess in the tension
of DNASGVS from NASGVS remains finite.
To see this, let $T(\tilde\omega ;L)$ be a logarithmically divergent mass 
with $L$ being a IR cut off scale in the unit of $\mu^{-1}$, 
\beq
\frac{T(\tilde\omega;L)}{2\pi v^2} &=& \int_0^L d\rho\ \rho\bigg[
\left(
h_1'{}^2 + h_2'{}^2
+ \frac{\left(h_1^2 (1-w_1)^2+h_2^2 w_1^2\right)}{\rho^2}
\right)
+ \frac{2 w_1'{}^2}{\rho^2}\non
&+& \frac{1}{4} \left\{\left(1-h_1^2\right)^2+\left(1-h_2^2\right)^2\right\}\non
&+& \frac{\tilde\omega^2}{2}\left\{
f'{}^2+\frac{f^2 (1-2 w_1)^2}{\rho^2}
+ \frac{1}{2}  \left(\left(f^2+1\right) \left(h_1^2+h_2^2\right) -4 f h_1 h_2 \right)
\right\}
\bigg].
\eeq
Then our guess is that 
$\delta T(\tilde \omega)$ defined by
\beq
\delta T(\tilde \omega) = T(\tilde \omega;L) - T(0;L),
\eeq
is finite.
We numerically compute $\delta T$ by changing the size of computational domain, 
and we find that $\delta T$ is indeed independent of size of computational box $L$ 
and is finite. 
The results are summarized in Table~\ref{tab:1}.

\begin{figure}[t]
\begin{center}
\includegraphics[width=13cm]{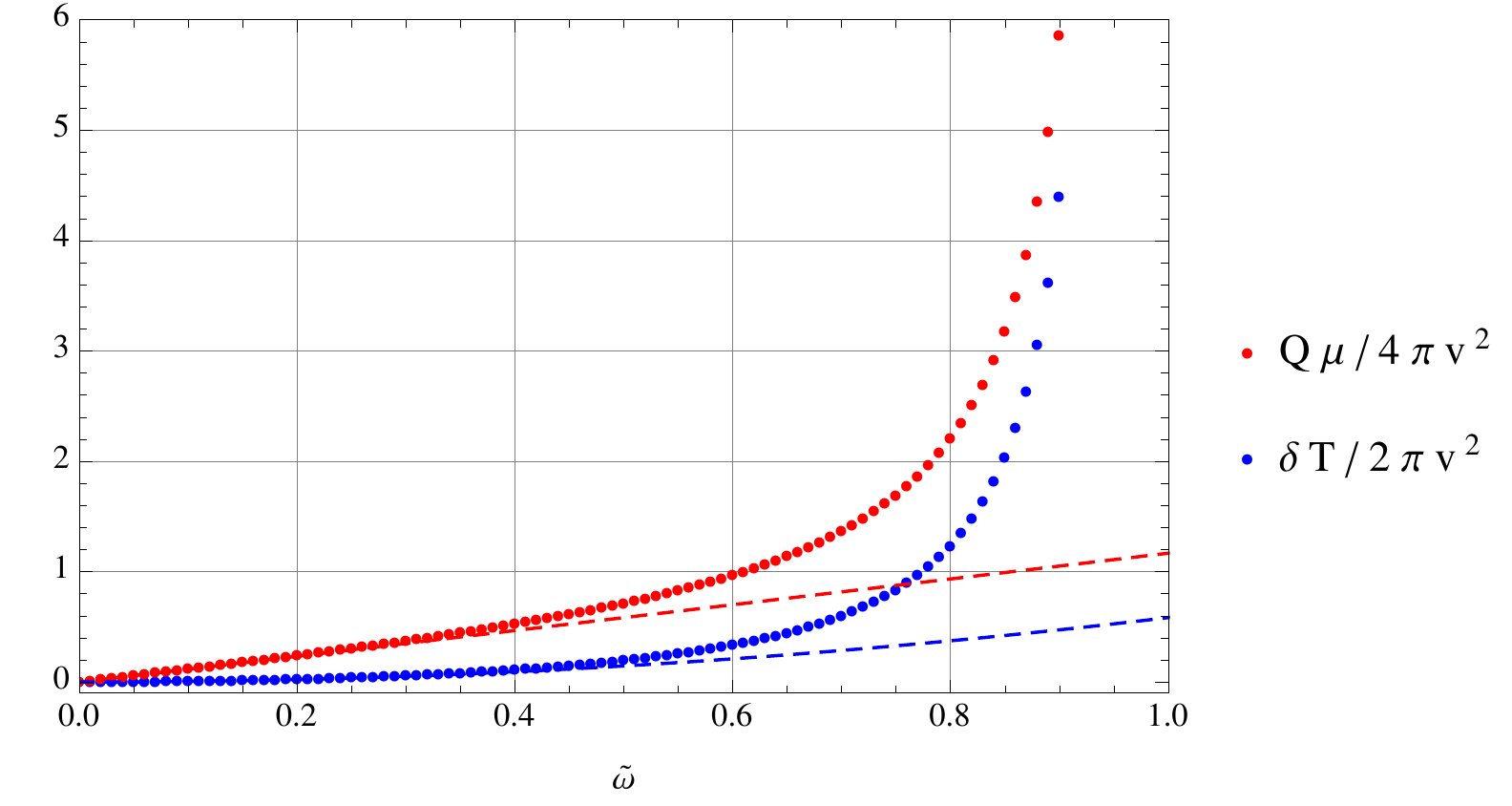}
\caption{Numerical results for $Q^3$ (red dots) and $\delta T$ (blue dots) for 
$0.1 \le \tilde \omega \le 0.9$. The asymptotic behaviors given in Eqs.~(\ref{eq:M_SGV})
and (\ref{eq:Q_SGV})
for small $\tilde \omega$
are shown in dashed blue and red lines, respectively.}
\label{fig:MQw_SGV}
\end{center}
\end{figure}

Behaviors of $Q^3$ and $\delta T$ as functions of $\tilde \omega$ $(0\le\tilde\omega\le0.90)$
are shown in Fig.~\ref{fig:MQw_SGV}.
For small $\tilde \omega$, as can be seen from Fig.~\ref{fig:MQw_SGV},
they are well approximated by linear and quadratic functions, respectively.
We find numerical fitting curves
\beq
\frac{\delta T(\tilde \omega)}{2\pi v^2} &=& \xi_1\, \tilde \omega^2 + {\cal O}\left(\tilde\omega^4\right),
\qquad \xi_1 = 0.584,
\label{eq:M_SGV}\\
\frac{\mu Q^3(\tilde \omega)}{2\pi v^2} &=& 2 \xi_2\, \tilde \omega + {\cal O}\left(\tilde\omega^2\right),
\qquad \xi_2 = 1.168.
\label{eq:Q_SGV}
\eeq
The coefficients $\xi_1$ and $\xi_2$ are slightly different from those in Eq.~(\ref{eq:num_MQ_U2}) for
the supersymmetric model. 
Nevertheless, a similar relation to the supersymmetric case holds
\beq
2\xi_1 = \xi_2.
\eeq
The reason why this special relation holds can be clearly explained from the view point of 
a low energy effective theory on the vortex world volume.
The orientational zero modes are associated with the spontaneous symmetry breaking
$SU(2)_{\rm C+F} \to U(1)_{\rm C+F}$, so that the corresponding effective Lagrangian 
should be a non-linear sigma model whose target space is 
$\mathbb{C}P^1 \simeq SU(2)_{\rm C+F}/U(1)_{\rm C+F}$
\beq
{\cal L}_{{\rm eff};SU(2)}^{(2)} 
= \tilde \beta \frac{\p_\alpha \phi \p^\alpha \phi^*}{\left(1+|\phi|^2\right)^2}.
\label{eq:eff_Lag_SU}
\eeq
Unlike the case of supersymmetric case, the overall coefficient $\tilde \beta$ cannot be analytically 
obtained but can be determined by numerical computations~\cite{Eto:2009bh}.
But here, for a while, we leave it an unknown constant.
As before, we are interested in a time-dependent solution $\phi = e^{i\omega t}$.
Then, we find excess in the tension and the
Noether charge per unit length
\beq
\delta T_{\rm eff} = \tilde \beta \frac{|\dot\phi|^2}{\left(1+|\phi|^2\right)^2} 
= \frac{\tilde \beta}{4}\omega^2,\qquad
q^3 = i \tilde \beta \frac{\phi\p_0\phi^* - \phi^*\p_0 \phi}{(1+|\phi|^2)^2}
= \frac{\tilde \beta}{2}\omega.
\label{eq:eff_SVG_MQ}
\eeq
Comparing this with Eqs.~(\ref{eq:M_SGV}) and (\ref{eq:Q_SGV}) and remembering the relation
$Q^3 = 2 q^3$, 
irrespective of value of $\tilde\beta$,
we reach the relation $2\xi_1 = \xi_2$.
Furthermore,  as a byproduct, we can determine the unknown constant $\tilde\beta$ 
by comparing Eqs.~(\ref{eq:M_SGV}) and (\ref{eq:eff_SVG_MQ})
\beq
\tilde \beta =  \frac{8\pi v^2 \xi_1}{\mu^2} = 2\xi_1\times\frac{4\pi}{g^2} = 1.168 \times \frac{4\pi}{g^2}.
\label{eq:tilde_beta_indirect}
\eeq

In order to check validity of the above indirect computation for $\tilde\beta$, 
let us calculate
$\tilde\beta$ in a direct way~\cite{Eto:2009bh}. 
To this end, we mimic the derivation for $\beta$ done in Sec.~\ref{sec:low energy}.
First, we prepare $x^\alpha$ ($\alpha=0,3$) dependent background configurations
\beq
H &=& U\left[
v\left(
\begin{array}{cc}
h_1 & 0 \\
0 & h_2
\end{array}
\right)
\right] U^{-1},\\
W_1 + i W_2 &=& U \left[
-i\frac{e^{i\theta}}{2r}
\left(
\begin{array}{cc}
2w_1-1 & 0 \\
0 & -2w_1+1
\end{array}
\right)
\right]
U^{-1},
\eeq
with $U$ given in Eq.~(\ref{eq:ori_H}). Here $h_1$, $h_2$ and $w_1$ are a static solution of
Eqs.~(\ref{eq:eomSU_1}) -- (\ref{eq:eomSU_3}), see Fig.~\ref{fig:conf_SGV}.
For $W_{0,3}$, we use the same Ansatz given in Eq.(\ref{eq:W0_eff}).
Plugging these into the Lagrangian (\ref{eq:Lag_SU}) and picking up the terms quadratic in $\p_\alpha$,
we find the effective Lagrangian given in Eq.~(\ref{eq:eff_Lag_SU}) as
\beq
{\cal L}_{{\rm eff};SU(2)}^{(2)} 
= \int dx^1dx^2\ \Tr
\left[-\frac{1}{g^2}\hat F_{i\alpha}\hat F^{i\alpha} + \hat\D_\alpha H (\hat\D^\alpha H)^\dagger\right]
= \tilde \beta \frac{\p_\alpha \phi \p^\alpha \phi^*}{\left(1+|\phi|^2\right)^2}.
\label{eq:derive_eff_Lag_SU(2)}
\eeq
with
\beq
\tilde\beta = \frac{4\pi}{g^2} \int d\rho\ \rho\left[
\lambda'{}^2 + \frac{(1-2w_1)^2}{\rho^2}(1-\lambda)^2
+ 
\frac{\lambda^2}{2} \left(h_1^2+h_2^2\right) +  (1-\lambda ) (h_2-h_1)^2
\right]
\label{eq:tilde_beta}
\eeq
\begin{figure}[t]
\begin{center}
\includegraphics[width=10cm]{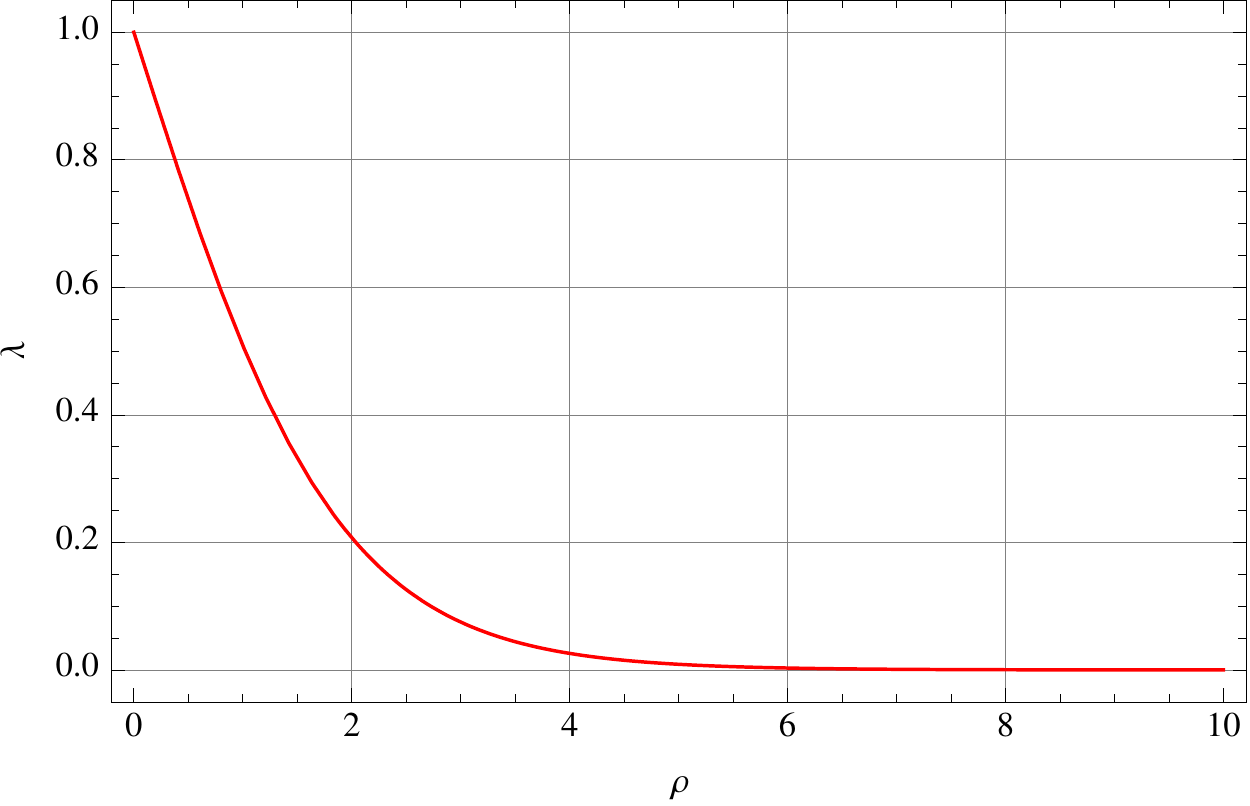}
\caption{A numerical solution of $\lambda$ for Eq.~(\ref{eq:eom_lambda}).}
\label{fig:lambda}
\end{center}
\end{figure}
This is slightly different from $\beta$ given in Eq.~(\ref{eq:beta}). If we replace $h_1 \to h$, $h_2 \to 1$
and $2w_1 \to w$,
$\beta$ and $\tilde \beta$ are formally identical.
However, since $h_1$, $h_2$ and $w_1$ do not solve the BPS equation
but solve the equations of motion (\ref{eq:eomSU_1}) -- (\ref{eq:eomSU_3}), the integral 
in Eq.~(\ref{eq:tilde_beta}) is not identical to $\beta = 4\pi/g^2$.
In order to find an appropriate $\lambda$, we need to numerically solve the equation of motion
\beq
\lambda ''+\frac{\lambda '}{\rho} +\frac{ (1-2w_1)^2}{\rho^2} (1-\lambda) + 
\frac{1}{2} \left((h_1^2+h_2^2)(1-\lambda) -2 h_1h_2\right) = 0.
\label{eq:eom_lambda}
\eeq
The boundary conditions are of the form
\beq
\lambda(0) = 1,\qquad
\lambda(\infty) = 0.
\eeq
A numerical solution is shown in Fig.~\ref{fig:lambda}. 
Plugging the numerical configurations of $h_1$, $h_2$, $w_1$ and $\lambda$ into Eq.~(\ref{eq:tilde_beta})
and integrating it over $\rho$, we get the following result
\beq
\tilde \beta = 
1.167 \times \frac{4\pi}{g^2}.
\eeq
The coefficient is in good agreement with the one in Eq.~(\ref{eq:tilde_beta_indirect}),
which ensures that the low energy effective theory gives a correct view point to 
understand DNASGVSs at least for small $\tilde \omega$.

\begin{figure}[t]
\begin{center}
\includegraphics[width=14cm]{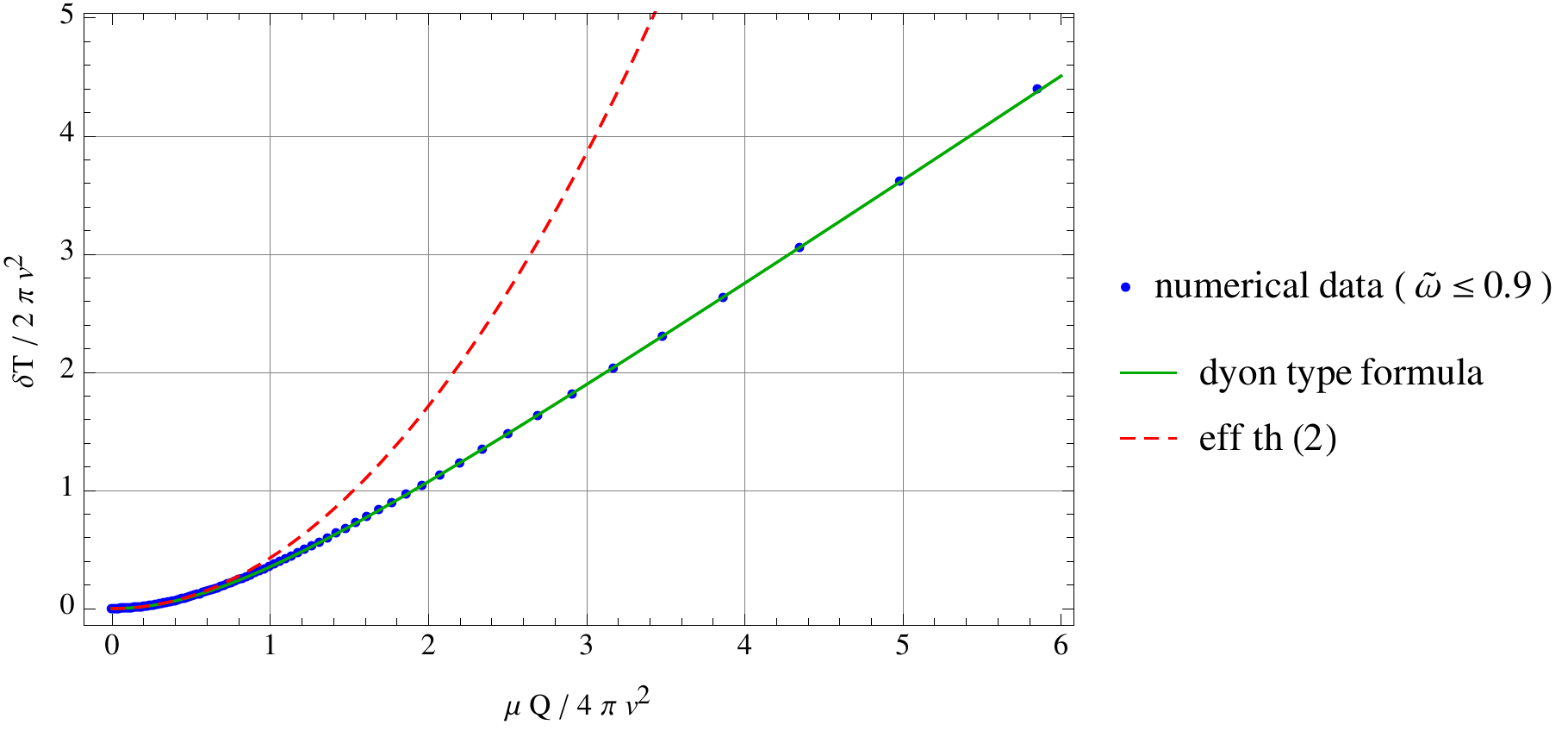}
\caption{Relation between $Q^3$ and $\delta T$ for a dyonic semi-global non-Abelian vortex.
The dots shows the numerical results, the red-dashed line corresponds to Eq.~(\ref{eq:eff_SVG_MQ}), and
the green solid line stands for the square root formula (\ref{eq:square_root_mass_SGV}).} 
\label{fig:QM_SU_k1}
\end{center}
\end{figure}

Finally, we turn to a global aspect including not only small $\tilde\omega$ but also
large $\tilde \omega \lesssim 1$ which cannot
be reproduced by the effective theory.
We plot $\delta T$ as a function of $Q^3$ with $\tilde \omega$ being a parameter in Fig.~\ref{fig:QM_SU_k1}.
The result is again surprising. Remember that 
the tension of NASGVS logarithmically diverges
and it is not BPS state even when $\tilde \omega = 0$. Nevertheless, the excess $\delta T$
obeys the square root formula quite well, which is  similar to the supersymmetric model
\beq
\delta T\big|_{k=1} 
\cong \sqrt{ (2\pi v^2\times 0.963)^2 + 0.806 \left(\frac{\mu Q^3}{2}\right)^2} - 2\pi v^2 \times 0.963.
\label{eq:square_root_mass_SGV}
\eeq
Here the second term in the left hand side is an artifact which is added 
for tuning $\delta T$ to be zero at $\tilde \omega = 0$.
Similarly, we find the following square root formulae for higher winding axially symmetric configurations
\beq
\delta T\big|_{k=2} &\approx& \sqrt{(2\pi v^2 \times 2.655)^2 + 0.770 \left(\frac{\mu Q^3}{2}\right)^2} - 2\pi v^2 \times 2.655,\\
\delta T\big|_{k=3} &\approx& \sqrt{(2\pi v^2 \times 5.392)^2 + 0.766 \left(\frac{\mu Q^3}{2}\right)^2} - 2\pi v^2 \times 5.392.
\eeq
Comparing these formulae with that for the supersymmetric model given in Eq.~(\ref{eq:SUSY_mass_formula}),
the coefficients of $(Q^3)^2$ in the square root are not unique for different $k$.

For $k=1$ DNASGVS, we also numerically find the following global relation between $\delta T$ and $\omega$
\beq
\delta T\big|_{k=1} \cong
2\pi v^2\left(
0.063\, \tilde \omega^2+\frac{0.463 \tilde \omega^2}{0.890 - \tilde \omega^2}
\right).
\eeq
Although this fitting curve reproduces the numerical data quite well, it has a pole at $\tilde \omega = \sqrt{0.890}
\simeq 0.943$. This seems to be awkward, but in practice it is not a problem. This is because 
it is hard to get appropriate numerical solutions for $\tilde \omega \gsim 0.9$ with the same reason 
explained below Eq.~(\ref{eq:relaxation}). Anyway, the fitting curve works quite well for $\tilde\omega < 0.9$.
Then, expanding this in $\tilde \omega$, we have
\beq
\delta T\big|_{k=1} &\cong& 2\pi v^2\left(\xi_1\, \tilde \omega^2 + 0.585\, \tilde\omega^4 + \cdots\right)\non
&=& \frac{\tilde \beta}{4} \left( \omega^2 + \frac{\omega^4}{\mu^2} + \cdots\right),
\label{eq:expand_DNASGVS}
\eeq
where we used the relation $\tilde \beta = 2\xi_1\beta$ and replaced $0.585/\xi_1$ by 1 
because of $\xi_1 \approx 0.584$.
The first term corresponds to the leading order contribution from the effective theory up to
quadratic derivatives, see Eq.~(\ref{eq:eff_SVG_MQ}). Then the second one should be identified with
quartic derivative corrections to the effective theory. 
In general, derivation of higher derivative corrections is not easy task. 
For example, only the quartic derivative
correction was obtained for the BPS NALVS in Ref.~\cite{Eto:2009bh}.  To make matters worse, 
NASGVS is non-BPS. Hence, we guess that derivation gets further complicated. Instead of involving 
such complicated computations, as is given in Eq.~(\ref{eq:fourth_derivative}),
here we assume the effective theory of the fourth order to be
the following form
\beq
{\cal L}_{\rm eff}^{(2+4)} = \tilde\beta\left[
\frac{|\p_\alpha\phi|^2}{(1+|\phi|^2)^2} + \frac{\tilde\gamma}{\mu^2}\frac{|\p_\alpha\phi\p^\alpha\phi|^2}{(1+|\phi|^2)^4}
\right].
\eeq
Then, increment in the tension for the configuration $\phi = e^{i\omega t}$ reads
\beq
\delta T_{\rm eff}^{(2+4)} = \frac{\tilde\beta}{4}\left(\omega^2 + \frac{3\tilde\gamma}{4\mu^2}\omega^4\right).
\eeq
Comparing this with Eq.~(\ref{eq:expand_DNASGVS}), we are able to make a prediction for the
coefficient of the four derivative term
\beq
\tilde \gamma \approx \frac{4}{3}.
\eeq
We leave verification of this prediction as a future problem. 
Furthermore, as is done in Sec.~\ref{sec:predict}, we may pursue deriving all order higher derivative corrections
to the effective theory for NASGVS.

\section{Conclusion}
\label{sec:conc}

We began this paper with making the conjecture that
all dyonic solitons of the dyonic-instanton (dyon) type are always BPS and their spatial co-dimensions are 
always even (odd).
The conjecture is true for all known dyonic solitons so far in the literature.
In order to examine further the conjecture, we have focused on
the dyonic extension of the non-Abelian vortex strings 
both in the supersymmetric and non-supersymmetric Yang-Mills-Higgs theories.

In the supersymmetric $U(\NC)$ Yang-Mills-Higgs model, the non-dyonic NALVSs
are the BPS states while their dyonic extensions, DNALVSs, can be either BPS or non-BPS
according to amount of the dyonic charges. This is a distinctive feature of DNALVS.
As expected from the conjecture, its BPS tension formula is
of the dyonic-instanton type. 
How about the non-BPS tension formula? The conjecture does not say anything about non-BPS states.
So we numerically solved the equations of motion and found the non-BPS tension formula.
Surprisingly, 
the non-BPS tension formula is approximately of the dyon-type.
For BPS states, the simple mass formulae can be understood from the perspective of
central charges of super algebra. In contrast, 
there are no reasons that such simple formulae like the dyonic-instanton or dyon type hold for non-BPS
states.  Thus, it still remains a question why the dyon-type formula for the non-BPS DNALVS holds, 
although it is the approximate relation.

Furthermore, we have studied a dyonic extension of 
another kind of non-Abelian vortex string, NASGVS, in the non-supersymmetric
Yang-Mills-Higgs model with $SU(\NC)$ gauge symmetry.
Regardless that NASGVS has a dyonic charge or not, it is always non-BPS, and its tension
is logarithmically dirvegent as a global vortex. Nevertheless, we found
that an increment in the tension due to an additional dyonic charge not only remains finite but also
approximately obeys the dyon-type tension formula.

Thus, we have found the common property for the different non-BPS dyonic vortex strings 
in the SUSY and non-SUSY Yang-Mills-Higgs theories.
Since the vortex strings are non-BPS and the property holds among SUSY and non-SUSY theories,
a reason for the property to hold has nothing to do with either BPS-ness or SUSY.
Or rather, we should make our attention on the common feature that the vortex strings have the 
normalizable non-Abelian zero modes. The dyonic extension can be understood as 
the rotation of the internal orientational modes. For the BPS case, the profile functions of 
the scalar fields and the gauge fields are  not affected 
by the additional motion of the internal orientations. 
Therefore, the tension formula is just the dyonic-instatnton type that the topological term
and the Noether charge are merely added.
On the contrary, for the non-BPS cases, the profile functions are deformed.
Thus, it is natural that the non-BPS tension formula changes from the dyonic-instanton type. 
What is not natural is that the formula is approximated very well by the dyon-type formula.

In order to have a deeper insight for understanding this, we have studied the low energy effective theory for
the internal orientation zero modes. While there are many works on the effective theory in the literature, 
much of them are on the lowest order theory. Namely, it includes only the quadratic 
derivative terms. There are also very few works for the higher derivative corrections 
of the quartic order~\cite{Eto:2012qda}.
However, such effective theories with lower order terms are not enough to understand the global
property of the non-BPS tension formula. In order to have a breakthrough, we have made the prediction
for the low energy effective theory (\ref{eq:all_order}) 
including the higher derivative corrections to the all order.
This has been done by combining the global tension formula (\ref{eq:global}) which has been obtained
by the numerical computation and the Ansatz (\ref{eq:eff_Lag_allorder}).
We have shown that the effective action has correct properties that the true effective theory has to have.
Furthermore, we have derived the implicit relation between the tension and the Noether charge
from Eqs.~(\ref{eq:all_tension}) and (\ref{eq:all_q}), which reproduces the numerical results and explain 
a reason why the mysterious dyon-type tension formula approximately holds.

Understanding the tension formula for the non-BPS dyonic strings is a practical purpose of this paper.
On the way to carry out the mission, we have found the efficient method to derive the low energy effective
theory with the all order higher derivative corrections. We expect that our indirect method can be 
applied also to other solitons like magnetic monopoles, instantons and so on.

Apart from things on the tension formula, we have also found the novel dyonic soliton, 
the spiral DNALVS, which is generated by boosting DNALVS along the string.
The orientation vector spins with angular frequency $m/\sqrt{1-u^2}$ while it twists 
around the string axis with wavenumber $mu/\sqrt{1-u^2}$ (the phase velocity
is $u$). We have confirmed that existence of the spiral DNALVS from the view points of the original 
Yang-Mill-Higgs theory in $3+1$ dimensions and the low energy effective theory in $1+1$ dimensions.

Before closing this section, let us make comments on  future directions.
Firstly, we would like to mention about possibility of detecting dyonic solitons in nature.
Although importance of topological solitons in high energy physics is widely accepted,
non of them have been detected in laboratories thus far.
No non-dyonic solitons have been found, much less dyonic solitons.
Nevertheless, we entertain hope that DNASGVS studied in Sec.~\ref{sec:massless_dyonic_SGV} would 
be found inside a compact star such as a quark star or optimistically
in a core of a neutron star. It is expected that QCD enters color superconducting phase at  a 
high baryon density region. Especially, at asymptotically high density region, the phase goes into 
so-called the CFL phase where QCD is weakly coupled theory. The effective non-Abelian Ginzburg-Landau
theory was obtained~\cite{Giannakis:2001wz,Iida:2002ev}, which is similar to the Lagrangian dealt with
in Sec.~\ref{sec:massless_dyonic_SGV}.
A possibility of finding non-dyonic NASGVSs in the CFL phase has been already pointed out~\cite{Balachandran:2005ev,Nakano:2007dr,Nakano:2008dc,Eto:2009kg,Eto:2009bh,Eto:2009tr,Eto:2011mk,Eto:2013hoa}.
So it would be natural to expect  that not only DNASGVS but also spiral DNASGVS exist. 
We will study DNASGV in the context of high density QCD elsewhere.
Secondly, the low energy effective Lagrangian including the all order derivative corrections given in
Eq.~(\ref{eq:all_order}) is bosonic. Since NALV is a half BPS state, the effective Lagrangian 
should be generalized to a supersymmetric form. For that purpose, it would be useful to make use of
a manifestly supersymmetric way for constructing higher derivative corrections which
was recently found~\cite{Nitta:2014fca}. 

\section*{Acknowledgements} 
This work is supported by Grant-in Aid for Scientific Research 
No.26800119 (M.\ E.). M.\ E. thanks to M.\ Arai, F.\ Blaschke and M.\ Nitta for useful comments.

\appendix

\section{BPS dyonic solitons}

\subsection{Dyonic instantons}

Dyonic instantons in $1+4$ dimensional $\N=1$ supersymmetric  $SU(2)$ Yang-Mills theory can be
found through the following Bogomol'nyi completion of Hamiltonian~\cite{Lambert:1999ua}
\beq
M_{\text{d-inst}} &=&  \int d^4x\ \Tr\left[ F_{0I}^2 +
\frac{1}{2}F_{IJ}^2 + (\D_0\Sigma)^2 + (\D_I\Sigma)^2 \right] \non
&=& \int d^4x\ \Tr\left[
\left(F_{0I} - \D_I\Sigma\right)^2 +
\frac{1}{4} \left(F_{IJ} - \tilde F_{IJ}\right)^2 + (\D_0\Sigma)^2 + \frac{1}{2} F_{IJ}\tilde F_{IJ}
+ 2 F_{0I}\D_I\Sigma
\right] \non
&\ge& \int d^4x\ \Tr\left[\frac{1}{2} F_{IJ}\tilde F_{IJ}\right] + \int d^4x\ \Tr\left[2 F_{0I}\D_I\Sigma\right],
\label{eq:Bogo_inst}
\eeq
with $I,J=1,2,3,4$, $F_{\mu\nu} = \p_\mu W_\nu - \p_\nu W_\mu + i g \left[W_\mu,W_\nu\right]$,
$\D_\mu \Sigma = \p_\mu + i g \left[W_\mu,\Sigma\right]$, and $\Tr[T^aT^b] = \delta^{ab}/2$. 
The energy bound is saturated when the BPS equations $F_{IJ} = \tilde F_{IJ}$, $F_{0I} = \D_I\Sigma$, $\D_0\Sigma=0$ and 
Gauss's law $\D_IF_{0I} = i\left[\Sigma,\D_0\Sigma\right] =0$ are satisfied.
The self-dual equations $F_{IJ} = \tilde F_{IJ}$ are identical to those for static instantons. 
A famous one-instanton solution in singular gauge is given by
$
W_I = - \frac{1}{g}\frac{\bar\eta_{IJ}x_J\rho^2}{x^2(x^2+\rho^2)},
$
where $\bar \eta_{IJ}$ is the anti self-dual 't Hooft tensor and $\rho$ corresponds to size modulus.
Plugging this into the first term in the last line of Eq.~(\ref{eq:Bogo_inst}), we have
$
Q_{\rm i} =  g^2 \int d^4x\ \Tr\left[\frac{1}{2} F_{IJ}\tilde F_{IJ}\right]  = 8\pi^2
$
(the gauge coupling has mass dimension $-1/2$).
The remaining two equations $F_{0I} = \D_I\Sigma$ and $\D_0\Sigma=0$ are solved by
$W_0 = -\Sigma$ which we assume time independent functions. Finally, we determine $\Sigma$ by
solving the Gauss's law $\D_I\D_I\Sigma = 0$. A solution is known to be
$
\Sigma = v \frac{x^2}{x^2+\rho^2} \frac{\sigma_3}{2}.
$
Here, $v$ is vacuum expectation value at spatial infinity. When $v\neq0$, 
the gauge symmetry is spontaneously 
broken to $U(1)$ subgroup. This gives rise to a non-zero electric charge which stabilizes 
the size of instanton by the relation
$
Q_{\rm e} = \int d^4x\ \p_I\Tr\left[2 F_{0I}\frac{\Sigma}{v}\right] 
= \int d^4x\ \Tr\left[2 F_{0I}\frac{\D_I \Sigma}{v}\right]
= 4\pi^2 \rho^2 v.
$
In summary, the one dyonic instanton with the electric charge $Q_{\rm e}$ has the BPS mass
\beq
M_{\text{d-inst}} = \frac{Q_{\rm i}}{g^2} + Q_{\rm e}v.
\eeq
Note that the spatial gauge field configurations $W_I$ are independent of $Q_{\rm e}$ and the BPS mass is
summation of the topological term $8\pi^2/g^2$ and electric contributions $Q_{\rm e}v$ by $W_0$ and $\Sigma$.

\subsection{Dyons}

Let us next recall the BPS dyons in $1+3$ dimensional $\N=2$ supersymmetric $SU(2)$ Yang-Mills theory.
The Bogomol'nyi completion for the Hamiltonian is performed in the following way
\beq
M_{\text{dyon}} &=& \int d^3x\ \Tr\left[
E_i^2 + B_i^2 + |\D_0\Sigma|^2 + |\D_i\Sigma|^2 
+ \frac{g^2}{4}\left[\Sigma,\Sigma^\dagger\right]^2
\right] \non
&=& \int d^3x\ \Tr\bigg[
\left(E_i -  \D_i\Sigma \sin\alpha \right)^2 + \left(B_i -  \D_i\Sigma \cos\alpha\right)^2
+ (\D_0\Sigma)^2 \non
&&+ 2  B_i \D_i\Sigma  \cos\alpha + 2  E_i \D_i \Sigma \sin\alpha
\bigg] \non
&\ge& \left(\int d^3x\ \Tr\left[2B_i\D_i\Sigma\right]\right) \cos\alpha + 
\left(\int d^3x\ \Tr\left[2E_i\D_i\Sigma\right]\right) \sin\alpha,
\label{eq:Bogo_mono}
\eeq
with
$B_i = \frac{1}{2}\epsilon_{ijk}F_{jk}$, $E_i = F_{0i}$, $i,j,k = 1,2,3$ and
$\alpha$ being an arbitrary constant. We set $\Sigma=\Sigma^\dagger$ for the second equality.
With the Bianchi identity $\D_i B_i = 0$ and Guass's low $\D_i E_i = ig\left[\Sigma,\D_0\Sigma\right] = 0$,
we have $Q_{\rm m} = \frac{2}{v} \int d^3x\ \p_i\Tr\left[B_i\Sigma\right] 
= \frac{2}{v}\int d^3x\ \Tr\left[B_i\D_i\Sigma\right]$ and 
$Q_{\rm e} = \frac{2}{v} \int d^3x\ \p_i\Tr\left[E_i\Sigma\right] 
= \frac{2}{v}\int d^3x\ \Tr\left[E_i\D_i\Sigma\right]$, where $v$ stands for vacuum expectation value of $\Sigma$
at spatial infinity. Plugging these into Eq.~(\ref{eq:Bogo_mono}),
the energy bound reduces to $M_{\rm dyon} \ge v(Q_{\rm m} \cos\alpha + Q_{\rm e} \sin\alpha)$.
Although $\alpha$ is arbitrary, the most stringent bound is obtained when $\tan\alpha = Q_{\rm e}/Q_{\rm m}$.
Thus BPS mass formula for the BPS dyon becomes
\beq
M_{\rm dyon} = v\sqrt{Q_{\rm m}^2 + Q_{\rm e}^2}.
\eeq
Note that the BPS equation $B_i = \frac{Q_{\rm m}}{\sqrt{Q_{\rm m}^2 + Q_{\rm e}^2}} \D_i \Sigma$
obviously depends on the electric charge $Q_{\rm e}$, so that the spatial gauge field configurations (the magnetic
fields) change from those for the magnetic monopole without electric charge.
Furthermore, the BPS mass formula is not mere superposition of the topological charge and the 
electric charge unlike the case of the dyonic instants.

\subsection{Q-lumps}

There are low dimensional analogue of the dyonic instantons and dyons. They are
Q-lumps in $1+2$ dimensions and Q-kinks in $1+1$ dimensions~\cite{Leese:1991hr}. 
One of the simplest model for the Q-lumps is $1+2$ dimensional
massive non-linear sigma model whose target space is $\mathbb{C}P^1$ manifold. 
Let $\phi$ be an inhomogeneous complex coordinate of $\mathbb{C}P^1$, then the Lagrangian is
given by
$
{\cal L} = \frac{\left|\p_\mu \phi\right|^2 - m^2 |\phi|^2}{\left(1+|\phi|^2\right)^2}.
$
The Hamiltonian can be cast into the following perfect square form 
\beq
M_{\text{Q-lump}} &=& \int d^2x\ \frac{1}{\left(1+|\phi|^2\right)^2}\left[
|\dot \phi|^2 + |\p_i\phi|^2 + m^2 |\phi|^2
\right] \non
&=& \int d^2x\ \frac{
|\bar \p \phi|^2 + |\dot\phi \mp i m \phi|^2
+ i \left(\p_1\phi\p_2\phi^* - \p_2\phi\p_1\phi^*\right)
\mp im \left(\dot\phi\phi^* - \phi\dot\phi^*\right)
}{\left(1+|\phi|^2\right)^2} \non
&\ge& \int d^2x\ \frac{
 i \left(\p_1\phi\p_2\phi^* - \p_2\phi\p_1\phi^*\right)
}{\left(1+|\phi|^2\right)^2}
\mp  m \int d^2x\ \frac{i\left(\dot\phi\phi^* - \phi\dot\phi^*\right)
}{\left(1+|\phi|^2\right)^2},
\label{eq:Bogo_lump}
\eeq
with $\bar \p = \p_1 + i \p_2$.
The BPS energy bound is saturated for the solutions of the BPS equations
$\bar\p \phi = 0$ and $\dot\phi = \mp i m \phi$. Generic solution is $\phi = e^{\mp i m t}f(z)$ for
an arbitrary rational function $f(z)$ of the complex coordinate $z= x+iy$.
The first term in the last line of Eq.~(\ref{eq:Bogo_lump}) is topological term
$Q_{\rm L} = \int d^2x\ \frac{
 i \left(\p_1\phi\p_2\phi^* - \p_2\phi\p_1\phi^*\right)
}{\left(1+|\phi|^2\right)^2} = 2\pi k$ with $k$ being positive integer.
The second term is nothing but the Noether charge for the $U(1)$ global transformation $\phi \to e^{i\alpha}\phi$,
$Q_{\rm N} = \int d^2x\ \frac{i\left(\dot\phi\phi^* - \phi\dot\phi^*\right)
}{\left(1+|\phi|^2\right)^2}$.
Therefore, the BPS mass for the Q-lump is similar to the one for the dyonic instantons\footnote{
The Noether charge $Q_{\rm N}$ is finite only for $k\ge 2$.}
\beq
M_{\text{Q-lump}} = Q_{\rm L} + m |Q_{\rm N}|.
\eeq
Note that one of the BPS equation $\bar\p\phi=0$ is same as that for the static BPS lumps for
the massless case with $m=0$. Therefore, the energy density contributed from spatial derivatives
are unaffected. Reflecting this fact, the BPS mass is mere summation of the topological mass and
Noether charge, which is parallel to the BPS dyonic instantons.

\subsection{Q-kinks}

The last example is BPS Q-kinks~\cite{Abraham:1992vb,Abraham:1992qv}. The simplest model is 
again the massive $\mathbb{C}P^1$ non-linear sigma model  in $1+1$ dimensions.
The Bogomol'nyi completion of Hamiltonian reads
\beq
M_{\text{Q-kink}} &=& \int dx\ \frac{1}{(1+|\phi|^2)^2}
\bigg[
|\dot\phi \mp i m (\sin \alpha) \phi|^2
+ |\phi' - m (\cos \alpha) \phi|^2 \non
&&\mp im\sin\alpha(\dot\phi \phi^* - \dot \phi^*\phi)
+ m \cos \alpha(\phi'\phi^*+\phi^*{}'\phi)
\bigg] \non
&\ge& 
m \cos\alpha \int dx\ \frac{\phi'\phi^*+\phi^*{}'\phi}{(1+|\phi|^2)^2}
\mp m \sin\alpha
\int dx\ \frac{i(\dot\phi \phi^* - \dot \phi^*\phi)}{(1+|\phi|^2)^2},
\label{eq:Bogo_kink}
\eeq
where $\alpha$ is an arbitrary constant, and dot and prime stand for derivatives by $t$ and $x$, respectively.
Inequality is saturated by solutions of the BPS equations
$
\phi' = m (\cos\alpha) \phi
$ and
$
\dot\phi = \pm i m (\sin \alpha) \phi
$.
The first integral in the second line of Eq.~(\ref{eq:Bogo_kink}) gives a topological charge
$
Q_{\rm k} = \int dx\ \frac{\phi'\phi^*+\phi^*{}'\phi}{(1+|\phi|^2)^2}
=  \left[ \frac{-1}{1+|\phi|^2}\right]^\infty_{-\infty} = 1,
$ while the second integral is again the Noether charge
$
Q_{\rm N} = \int dx\  \frac{-i(\dot\phi\phi^*-\dot\phi^*\phi)}{(1+|\phi|^2)^2}.
$
Since $\alpha$ is arbitrary, for given $Q_{\rm k}$ and $Q_{\rm N}$, the most stringent bound
is obtained for $\tan\alpha = Q_{\rm N}/Q_{\rm k}$. The BPS Q-kink solution is given by
$
\phi = 
\exp\left(
\frac{Q_{\rm k}}{\sqrt{Q_{\rm k}^2 + Q_{\rm N}^2}} m(x-x_0)
+ \frac{iQ_{\rm N}}{\sqrt{Q_{\rm k}^2 + Q_{\rm N}^2} }m t
\right)
$.
The BPS mass of the Q-kink is of the dyon type
\beq
M_{\rm k} = m \sqrt{Q_{\rm k}^2 + Q_{\rm N}^2}.
\eeq
Note that, contrary to the Q-lumps, contribution to the energy density by the spatial kinetic term
does depend on the
Noether charge $Q_{\rm N}$. 
At the same time, the BPS mass formula is not mere sum of the topological term and
the Noether charge but the square root of sum of their squares.

\end{document}